\documentclass[12pt]{article}

\usepackage{color}
\usepackage{amsmath}
\usepackage{amssymb}
\usepackage{graphicx}
\usepackage{dcolumn}
\usepackage{bm}
\usepackage{hyperref}
\usepackage{nameref}
\usepackage{multirow,ulem}
\usepackage{array}
\usepackage{ulem}
\usepackage{authblk}
\usepackage{nameref}
\usepackage{setspace}
\usepackage{algorithm}
\usepackage[noend]{algpseudocode}
\usepackage{booktabs}
\usepackage{siunitx}
\usepackage{microtype}
\usepackage{threeparttable}
\newcommand{\keywords}[1]{\par\noindent\textbf{Keywords: }#1}
\newcommand{\jel}[1]{\par\noindent\textbf{JEL classification codes: }#1}

\usepackage[margin=1in]{geometry}

\newcolumntype{P}[1]{>{\centering\arraybackslash}p{#1}}

\title{Business cycle synchronization between the EU and Western Balkan candidate economies: A Wavelet Analysis}

\author{Petar Jolakoski\textsuperscript{1}, Viktor Stojkoski\textsuperscript{2,3}, Dragan Tevdovski\textsuperscript{2}}

\affil{\footnotesize
\textsuperscript{1}\footnotesize Research Center for Computer Science and Information Technologies, Macedonian Academy of Sciences and Arts, 1000 Skopje, N. Macedonia \\
\textsuperscript{2}\footnotesize Faculty of Economics, Ss.~Cyril and Methodius University in Skopje, 1000 Skopje, N. Macedonia \\
\textsuperscript{3}\footnotesize Center for Collective Learning, CIAS, Corvinus University of Budapest, Budapest, Hungary}

\date{March 30, 2026}

\begin{document}

\newtheorem{theorem}{Theorem}
\newtheorem{definition}{Definition}
\newtheorem{lemma}{Lemma}
\newtheorem{proposition}{Proposition}
\newtheorem{remark}{Remark}
\newtheorem{corollary}{Corollary}
\newtheorem{example}{Example}

\maketitle

\begin{abstract}
Business cycle synchronization between EU and Western Balkan candidate economies is usually modeled with aggregate time-domain correlations that mix short-run and long-run dynamics. This paper addresses that limitation by combining wavelet-based time-frequency decomposition with Bayesian zero-inflated beta regression. Using annual dyad-year data for 2001--2021, we estimate synchronization separately at shorter (1.5--4.5 years) and longer (4.5--8.5 years) horizons and relate each horizon to its correlates. The results show that EU--WB dyads are less synchronized than EU--EU dyads in the short run, and that trade deepening over time is more positively associated with short-run synchronization in EU--WB pairs. At longer horizons, the positive association between shared EU/EMU membership and synchronization weakens or reverses when the same country pair moves into deeper institutional integration, while differences across country pairs in average EU/EMU status become negligible. Over the same horizon, trade deepening within a pair is more consistently associated with synchronization, and more persistent structural dissimilarity is associated with lower synchronization. EU--WB dyads are no longer clearly less synchronized at these frequencies, and the remaining convergence pattern is more consistent with sectoral differences narrowing over time than with trade. These findings indicate that synchronization channels are horizon-dependent and that conclusions based on single-horizon correlation measures can obscure the distinction between short-term coupling and structural convergence.
\end{abstract}


\jel{E32, C23, C11, F15, F41}

\section{Introduction}

Western Balkan (WB) economies view EU accession as the main strategic anchor for long-run institutional and economic convergence, but entry is conditional on meeting a demanding set of political, legal, and economic requirements. In particular, the EU's Copenhagen criteria emphasize (among other pillars) the existence of a functioning market economy and the capacity to cope with competitive pressure and market forces within the Union. These benchmarks implicitly require macroeconomic stability and resilience as integration deepens. In parallel, current EU policy toward the region places renewed emphasis on accelerating economic convergence and closer integration with the Single Market, underscoring that convergence remains incomplete and uneven across partners. To this end, the degree of business-cycle synchronization with the EU has become a natural diagnostic of real integration. This captures whether shocks and recoveries are increasingly shared, leading to stronger trade integration that can raise cyclical comovement. However, this also leaves room for divergence when specialization patterns and financial linkages differ across countries.

But to which extent are the Western Balkan economies' business cycles synced with the EU?

The research so far has focused mostly on time-domain measures of WB--EU synchronization, typically using contemporaneous or rolling correlations of output and related indicators, and it generally finds rising co-movement since the early 2000s with substantial cross-country heterogeneity~\cite{Gouveia2014,hildebrandt2015business}. In the WB evidence, trade emerges as the main positive synchronization channel, whereas FDI, bank flows, and remittances are associated with weaker synchronization~\cite{hildebrandt2015business}. These time-domain approaches, however, do not separately identify short-run business-cycle frequencies and longer-run structural co-movement, so higher co-movement may reflect temporary coupling rather than durable convergence. Similarly, evidence from the broader post-transition literature highlights substantial cross-country heterogeneity in synchronization with the euro area and shows that the estimated degree of co-movement is sensitive to data choices and empirical methods~\cite{fidrmuc2006meta,darvas2008business}. Beyond trade intensity, the literature also emphasizes that structural similarity and specialization patterns shape cyclical co-movement, while the role of other determinants, including financial and policy-related factors, remains more mixed across studies~\cite{imbs2004trade,de2008will,darvas2008business}. More broadly, this literature suggests that trade exposure alone does not guarantee durable synchronization, because cross-country differences in production structure and other transmission mechanisms also matter~\cite{imbs2004trade,de2008will}.

To bridge this gap, we apply a wavelet-based determinants analysis of business cycle synchronization between EU countries and three Western Balkan candidates: Montenegro, N. Macedonia, and Serbia.\footnote{We exclude Albania and Bosnia and Herzegovina from the analysis due do time-series limitations.} This approach allows us to decompose the synchronization and construct annual estimates of dyadic (country by country) short-run (1.5--4.5 years) and long-run (4.5--8.5 years) synchronization. We use these estimates to understand the potential macroeconomic factors that might drive the degree of synchronization. More generally, wavelet methods provide a flexible framework for analyzing economic time series jointly in the time and frequency domains, and several surveys have highlighted their usefulness for economists and finance researchers~\cite{crowley2007guide,schleicher2002introduction,genccay2002introduction}. Related methodological contributions show how these tools can be used to measure business-cycle variation across scales, construct practical time--frequency diagnostics, and move beyond simple bivariate settings~\cite{yogo2008measuring,torrence1998practical,grinsted2004application,aguiar2014continuous}. Beyond business-cycle applications, wavelet methods have also been used in economics and finance to study monetary-policy transmission, international stock-market comovement, and market risk~\cite{aguiar2008using,rua2009international,rua2012wavelet}.

We find that Western Balkan countries exhibit lower short-run synchronization compared to EU-EU pairs, while trade deepening is more positively associated with short-run synchronization in EU--WB dyads than in EU--EU pairs. At long-run frequencies, the average EU--WB synchronization gap largely disappears, and the remaining convergence pattern is associated more with sectoral differences narrowing over time within the same EU--WB dyad than with trade. These frequency-dependent convergence channels broadly align with the results observed for the full sample, where trade and structural similarity operate through different mechanisms at different time horizons.

\section{Western Balkans Road to Business Cycle Integration with the EU}

The European monetary integration relies fundamentally on the premise that sharing a common currency delivers net benefits when member economies fluctuate together, allowing a single monetary policy to stabilize the entire region. This premise is rooted in Optimum Currency Area (OCA) theory. In the classic formulation, Mundell emphasized adjustment to asymmetric shocks through factor mobility~\cite{mundell1961theory}, McKinnon stressed the openness criterion~\cite{mckinnon1963optimum}, and Kenen highlighted the stabilizing role of diversified production structures~\cite{kenen1969optimum}. Business cycle synchronization lies at the heart of this framework. If member states experience booms and recessions simultaneously, a common interest rate set by a central bank will be appropriate for all. Conversely, if cycles are asynchronous, a single monetary policy becomes destabilizing, overheating economies in expansion while deepening recessions in lagging members. Understanding what drives synchronization is therefore essential for evaluating the viability of WB countries to join the EU.

The determinants of synchronization are characterized by a fundamental theoretical ambiguity between two competing hypotheses: The Endogeneity Hypothesis~\cite{frankel1998endogenity} and the Specialization Hypothesis, associated with Krugman~\cite{Krugman1993}. The Endogeneity Hypothesis, posits that the criteria for an OCA are endogenous: the very act of integrating, like reducing trade barriers, forming a currency union, sets in motion forces that synchronize business cycles. Under this view, closer trade linkages transmit aggregate demand shocks across borders. For example, when the German economy expands, its demand for French imports rises, pulling the French economy up with it. This mechanism suggests that trade integration creates the conditions for higher synchronization, thus historical correlations are poor predictors of post-integration co-movement. Initial empirical studies generally reported a positive relationship between trade intensity and cycle correlation~\cite{de2008will,calderon2007trade}, although more careful re-examinations suggest that this effect is smaller and less robust than early estimates implied~\cite{inklaar2008trade}. A related refinement is that the composition of trade matters as much as its aggregate volume: integration that increases intra-industry trade tends to support business-cycle convergence, whereas integration that fosters inter-industry specialization may instead raise exposure to asymmetric shocks~\cite{fidrmuc2004endogeneity}. Related evidence from the broader European literature also suggests that the trade--synchronization relationship depends on sectoral similarity and the composition of trade, and that the marginal synchronizing effect of deeper integration may weaken over time~\cite{gianelle2017interregional,mendoncca2011shrinking,steinert2025endogeneity}. At the same time, deeper integration appears to synchronize demand-side disturbances more readily than supply-side shocks, implying that its effect depends on the type of shock under consideration~\cite{babetskii2005trade}.

The Specialization Hypothesis, by contrast, argues that economic integration allows firms to exploit economies of scale by concentrating production in specific regions. For example, this would explain the automobile manufacturing clusters in Central Europe to utilize skilled labor and supply chains (because of their proximity to the main manufacturers). Similarly, this hypothesis explains why tourism concentrates in the Mediterranean (to exploit natural endowments). This specialization leads to inter-industry trade rather than intra-industry trade. The critical implication is that as regions become more specialized, they become more vulnerable to asymmetric, sector-specific shocks. Consequently, Krugman predicted that deeper integration would lead to less synchronized business cycles over time, making monetary union potentially problematic.

However, these hypotheses are not necessarily contradictory, because they can operate at different horizons. Trade integration can raise short-run co-movement by strengthening demand spillovers and accelerating the transmission of common shocks, consistent with the endogeneity view. Over longer horizons, integration may also induce structural specialization and sectoral concentration, increasing exposure to asymmetric industry shocks and potentially weakening synchronization, as emphasized by the specialization hypothesis. Crisis-era evidence is also consistent with this view: co-movement intensified during the global financial crisis, yet synchronization remained heterogeneous across countries and business-cycle horizons~\cite{bekiros2015business}. Related evidence emphasizes that international business cycles reflect a combination of global, regional, sectoral, and country-specific shocks, so stronger comovement during common disturbances can coexist with persistent cross-country heterogeneity~\cite{kose2003international,karadimitropoulou2018advanced}.

So, what do these theories tell us about the WBs business cycle synchronization with the EU?

The literature on EU--WB synchronization~\cite{Gouveia2014,hildebrandt2015business} relies on time-domain measures, including correlation coefficients, concordance indices, and period-by-period correlation measures. These measures do not distinguish whether co-movement represents structural alignment or short-term coupling driven by supply-chain dynamics. Hildebrandt~\cite{hildebrandt2015business} identified trade as the primary positive driver of WB--EU synchronization, but found that FDI, bank flows, and remittances were associated with weaker synchronization. The key distinction is that these flows are described as procyclical with respect to the recipient WB economy, meaning that they tend to rise when the local economy is already expanding and fall when it weakens; this can amplify country-specific fluctuations rather than make them move more closely with the EU cycle. Without frequency decomposition, however, it remains unclear at which horizons this decoupling operates. Country-specific evidence for the Western Balkans remains heterogeneous. Serbia appears weakly synchronized with the euro area and sensitive to filtering choices~\cite{obradovic2013synchronization}. By contrast, N. Macedonia has been found to exhibit strong and rising synchronization with the Euro-zone, with foreign shocks transmitted quickly and persistently into domestic activity~\cite{filipovski2018business}. For Montenegro, direct synchronization evidence is thinner, but the euroization literature suggests that imported ECB policy has been broadly countercyclical for output even though domestic monetary transmission remains heavily mediated by local banking-sector conditions~\cite{sokic2013euroization}. Broader cross-country comparisons indicate that non-EU countries, including the Western Balkan candidates, are generally less synchronized with EU and euro-area cycles than the old EU core, though data constraints, such as the need to proxy missing quarterly GDP with industrial production for Montenegro and Bosnia-Herzegovina, limit comparability~\cite{kovavcic2017assessing}. In addition, recent IMF analysis~\cite{ma2025impact}, using a Bayesian GVAR model, documents that N. Macedonia experiences the largest spillovers from euro area real activity shocks among Western Balkan countries, roughly twice those of Serbia or Bosnia-Herzegovina, driven by its integration into German automotive supply chains. Whether this pronounced sensitivity reflects genuine structural convergence or a narrower dependency on German production cycles cannot be resolved by impulse response analysis alone, motivating the frequency-domain decomposition used here.

Wavelet evidence for the EU indicates that synchronization is both time-varying and frequency-specific, so country pairs may be synchronized at some horizons but not at others \cite{soares2011business}. This motivates separating short-run co-movement from long-run convergence in our analysis. Evidence since the introduction of the euro shows that co-movement increased overall, but important differences across groups of countries remained, so Europe still does not move as a single business cycle \cite{fidrmuc2018business,camacho2008european}. Moreover, evidence from post-transition CESEE economies is directly relevant for the Western Balkans: synchronization with the euro area is uneven across countries and appears to strengthen more clearly for some integration paths than for others~\cite{darvas2008business}. Within that broader pattern, stronger trade integration is generally associated with closer alignment, especially in less-integrated SEE economies~\cite{botric2016bilateral}, while the depth of EU integration shapes which channels of synchronization matter across country groups~\cite{botric2016exploring,botric2019business}.

Conventional time-domain approaches may miss important differences in the sign and amplitude of output gaps and can therefore give an incomplete picture of coherence~\cite{mink2012measuring}. They are also limited in tracking how co-movement evolves across frequencies and over time, which is precisely the dimension highlighted by time--frequency approaches~\cite{crowley2005decomposing}. Related methodological work shows that business-cycle co-movement can be characterized in different ways: turning-point approaches identify expansions and recessions~\cite{harding2002dissecting}, frequency-domain measures distinguish short-run from long-run co-movement across frequency bands~\cite{croux2001measure}, and factor models separate world, regional, and country-specific sources of fluctuations~\cite{kose2003international}. However, these approaches do not jointly show how synchronization varies across both time and frequency. Wavelet-based frequency decomposition addresses that gap by representing co-movement in the time--frequency domain and estimating synchronization separately across frequency bands. Aguiar-Conraria and Soares~\cite{soares2011business} use wavelet-based distance measures and cross-wavelet phase analysis to show that business-cycle synchronization in Europe is time- and frequency-dependent, with France and Germany forming the core, while Portugal, Greece, Ireland, and Finland do not display statistically relevant synchronization with Europe. Additional wavelet evidence reinforces this point: euro-area co-movement differs across frequency bands and over time, and a persistent core--periphery pattern appears across horizons~\cite{crowley2005decomposing}, while synchronization between CEECs and the EU-15 varies systematically across both time and frequency~\cite{njegic2017business}. Using dynamic correlation analysis rather than wavelets, Fidrmuc~\cite{fidrmuc2013china} examined the China--G7 relationship and found that synchronization was weak or negative at conventional business-cycle horizons, with trade integration not necessarily increasing co-movement. Taken together, these results suggest that cross-border integration may reinforce short-run transmission while leaving longer-run structural dynamics more weakly aligned.

The remainder of this paper is organized as follows. Section~\ref{sec:data} describes the data collection and variable construction process. Section~\ref{sec:methodology} develops the methodological strategy by deriving the wavelet coherence measure of synchronization and specifying an econometric model suited to our data. Section~\ref{sec:results} presents the results, beginning with wavelet evidence of synchronization and followed by estimates of the determinants of short-run and long-run synchronization for the full sample, with a dedicated subsection on Western Balkans convergence. Section~\ref{sec:conclusion} concludes and discusses the findings.

\section{Data}\label{sec:data}

We start by collecting monthly Industrial Production Index (IPI) data for three Western Balkans economies (N. Macedonia, Serbia, and Montenegro) and all EU member states from the IMF database (non-seasonally adjusted, January 2000--December 2023). IPI is a high-frequency measure of economic activity, making it suitable for measuring the business cycle synchronization between two economies.

We use these data to construct an annual measure of in-phase synchronization, capturing the degree to which two countries' IPI series move together at specific frequencies (using the wavelet coherence approach described in Section~\ref{sec:sync_dv}). In particular, we construct annual dyad-year panels by aggregating monthly wavelet coherence within two frequency bands: short-run (1.5--4.5 years) and long-run (4.5--8.5 years).

To understand the variation in synchronization we collect annual data on bilateral trade intensity, institutional integration indicators, structural similarity, fiscal and macro gaps, and financial depth measures, following \cite{ductor2016dynamics}. The sources and definitions of all variables are described in Table~\ref{tab:variables}.

We measure trade integration using two complementary dyad-year indicators. First, trade intensity captures the strength of bilateral goods-market linkages by scaling total two-way trade between countries $a$ and $b$ by their combined economic size. Second, financial openness proxies the depth of cross-border financial exposure by combining each country’s external assets and liabilities relative to GDP, summed at the dyad level. Together, these measures distinguish real integration through trade flows from financial integration through international balance-sheet linkages.

Institutional integration enters the model through mutually exclusive dyad-year dummies based on time-varying membership calendars. The construction proceeds in steps shown in Table~\ref{tab:variables}: first we define country-level membership indicators, then multiply them to create pair-level indicators, and finally construct two non-overlapping dummies for regression analysis. Each pair-year falls into exactly one of three categories: (i) not both EU (baseline), (ii) both EU but not both EMU, or (iii) both EMU. This allows us to separately identify single-market effects (EU-only) from monetary union effects (EMU).

Fiscal and macro differences are captured through dyad-year gaps that measure the extent to which countries face dissimilar nominal conditions and policy stances. We proxy these differences using the absolute gap in government expenditure (as a share of GDP) and the absolute gap in inflation, with larger gaps interpreted as greater macroeconomic divergence.

Structural factors account for persistent cross-country differences in production structures and development characteristics that shape how shocks transmit and how similarly economies respond. We include a measure of specialization distance based on differences in sectoral value-added shares, alongside gaps in capital stock per capita and urbanization, which capture broader differences in productive capacity and economic structure.

Financial depth is measured using alternative dyad-year gaps in domestic financial intermediation capacity. Specifically, we consider differences in liquid liabilities, financial system deposits, and bank deposits (each expressed relative to GDP).

Finally, other channels are captured through the remittances gap, measured as the absolute difference in personal remittances received (as a share of GDP). This variable proxies differences in reliance on external household income flows, which can affect domestic demand dynamics and the transmission of foreign shocks in ways that matter for business-cycle co-movement. Importantly, this measure is not equivalent to the remittance variables used in the literature, which are constructed as bilateral proxies based on total EU remittances to each Western Balkan country and the distribution of migrants across EU host countries. Instead, the measure used here captures cross-country differences in aggregate remittance dependence rather than bilateral remittance linkages.

To ensure data quality, we drop dyad-years with insufficient monthly coverage (fewer than 9 total months or fewer than 6 eligible months outside the cone of influence i.e. outside the edge region of the wavelet transform where estimates are less reliable), removing spurious zeros from low coverage while retaining structural zeros. For regressors, we construct derived country-level measures (including real capital per capita, log human capital, and external-balance ratios) before forming dyadic variables. Missing covariate values are filled by within-country linear interpolation for internal gaps of up to five years, without extrapolation beyond observed endpoints; we then build undirected dyad-year gaps as absolute cross-country differences. For more details see Appendix~\ref{app:data}.

Altogether, our final sample contains 7,473 dyad-year observations for 26 EU member states\footnote{Cyprus is excluded because complete covariate coverage is not available in the final dataset, especially for the financial variables.} and Western Balkan candidates (N. Macedonia, Serbia, Montenegro), covering the period between 2001 and 2021. The effective estimation sample is smaller in the regression models because of lag construction and related specification-specific restrictions; the corresponding model-specific observation counts are reported in the table notes.

\begin{table}[htbp]
\centering
\caption{Independent variable definitions}
\label{tab:variables}
\scriptsize
\begin{threeparttable}
\begin{tabular}{p{2.5cm}p{4.5cm}p{3.8cm}p{3.2cm}}
\toprule
\textbf{Variable} & \textbf{Definition} & \textbf{Interpretation} & \textbf{Source(s)} \\
\midrule
\multicolumn{4}{l}{\textit{Trade \& Integration}} \\
Trade intensity & $T_{ab,t} = \dfrac{E_{a\to b,t} + I_{a\leftarrow b,t}}{GDP_{a,t} + GDP_{b,t}}$ & Bilateral trade ($E$: exports, $I$: imports) relative to joint GDP & BACI HS92 (\cite{gaulier2010baci});\\[0.3cm]
Financial openness & $F_{ab,t} = \dfrac{A_{a,t}+L_{a,t}}{GDP_{a,t}} + \dfrac{A_{b,t}+L_{b,t}}{GDP_{b,t}}$ & Sum of external assets ($A$) and liabilities ($L$) relative to GDP for both countries (Lane \& Milesi-Ferretti IFI) & EWN Database ($A$: total assets excl.\ gold; $L$: total liabilities) \\[0.3cm]
\midrule
\multicolumn{4}{l}{\textit{Institutional Integration}} \\
Membership indicators & $M^{\mathrm{EU}}_{ct} = \mathrm{I}(c \in \mathrm{EU}_t)$, \quad $M^{\mathrm{EMU}}_{ct} = \mathrm{I}(c \in \mathrm{EMU}_t)$ & Indicator $\mathrm{I}(\cdot)$ returns 1 if country $c$ is member at $t$, 0 otherwise & Authors' construction from official accession calendars \\[0.2cm]
Pair indicators & $B^{\mathrm{EU}}_{gt} \equiv M^{\mathrm{EU}}_{at}M^{\mathrm{EU}}_{bt}$, \quad $B^{\mathrm{EMU}}_{gt} \equiv M^{\mathrm{EMU}}_{at}M^{\mathrm{EMU}}_{bt}$ & Equals 1 if both countries in pair $g=(a,b)$ are members at $t$ & Derived \\[0.2cm]
Both EU non-EMU & $D^{\mathrm{EU}\setminus\mathrm{EMU}}_{gt} = B^{\mathrm{EU}}_{gt}(1 - B^{\mathrm{EMU}}_{gt})$ & Both in EU but not both in euro area (regression variable) & Derived \\[0.2cm]
Both EMU & $D^{\mathrm{EMU}}_{gt} = B^{\mathrm{EMU}}_{gt}$ & Both in euro area (regression variable) & Derived \\[0.2cm]
\midrule
\multicolumn{4}{l}{\textit{Fiscal \& Macro Gaps}} \\
Government expenditure gap & $\Delta GE_{ab,t} = |GE_{a,t} - GE_{b,t}|$ & Absolute difference in government expenditure (\% of GDP) & World Bank WDI (NE.CON.GOVT.ZS) \\[0.2cm]
Inflation gap & $\Delta \pi_{ab,t} = |\pi_{a,t} - \pi_{b,t}|$ & Absolute difference in annual CPI inflation rates (\%) & World Bank WDI (FP.CPI.TOTL.ZG) \\[0.2cm]
\midrule
\multicolumn{4}{l}{\textit{Structural Factors}} \\
Specialization distance & $S_{ab,t} = \displaystyle\sum_{k=1}^{K} |s_{k,a,t} - s_{k,b,t}|$ & Sectoral dissimilarity ($s_k$: GDP share of sector $k$). Range: [0,2] when $K=3$ & World Bank WDI (NV.AGR/IND/SRV.TOTL.ZS) \\[0.3cm]
Capital services gap & $\Delta K_{ab,t} = |K_{a,t} - K_{b,t}|$ & Absolute difference in real capital services per capita & PWT 10.01 (rkna, pop) \\[0.2cm]
Urbanization gap & $\Delta U_{ab,t} = |U_{a,t} - U_{b,t}|$ & Absolute difference in urban population share (\%) & World Bank WDI (SP.URB.TOTL.IN.ZS) \\[0.2cm]
\midrule
\multicolumn{4}{l}{\textit{Financial Depth Proxies}} \\
Liquid liabilities gap & $\Delta LL_{ab,t} = |LL_{a,t} - LL_{b,t}|$ & Absolute difference in liquid liabilities (\% of GDP) & World Bank GFDD (GFDD.DI.05) \\[0.2cm]
Financial system deposits gap & $\Delta FSD_{ab,t} = |FSD_{a,t} - FSD_{b,t}|$ & Absolute difference in financial system deposits (\% of GDP) & World Bank GFDD (GFDD.DI.08) \\[0.2cm]
Bank deposits gap & $\Delta BD_{ab,t} = |BD_{a,t} - BD_{b,t}|$ & Absolute difference in bank deposits (\% of GDP) & World Bank GFDD (GFDD.OI.02) \\[0.2cm]
\midrule
\multicolumn{4}{l}{\textit{Other}} \\
Remittances gap & $\Delta R_{ab,t} = |R_{a,t} - R_{b,t}|$ & Absolute difference in personal remittances received (\% of GDP) & World Bank WDI (BX.TRF.PWKR.DT.GD.ZS) \\[0.2cm]
\bottomrule
\end{tabular}
\begin{tablenotes}
\footnotesize
\item \textit{Notes:} Subscripts $a$, $b$ denote the two countries in dyad $g=(a,b)$; $t$ denotes year. GFDD indicators are sourced from IMF \textit{International Financial Statistics} (IFS) as compiled by the World Bank Global Financial Development Database.
\end{tablenotes}
\end{threeparttable}
\end{table}

\section{Methodology}\label{sec:methodology}

Traditional measures of business cycle synchronization, for example Pearson correlations, rolling-window correlations, or HP-filtered series, mix distinct economic phenomena operating at different frequencies. This shortcoming poses an opportunity to analyze the contradictory conclusions from the literature about the macroeconomic determinants of business cycles syncronization.

We address this ambiguity through a four-stage empirical strategy. First, we apply continuous wavelet transforms to monthly industrial production indices, decomposing each time series into a time--frequency plane that reveals when and at which periodicities co-movement occurs. Second, we compute wavelet coherence, that is a localized measure of correlation in the time--frequency domain, and band-average it within short-run (1.5--4.5 years) and long-run (4.5--8.5 years) frequency bands. Third, we aggregate monthly coherence to annual dyad-year observations, constructing an in-phase synchronization index that captures both the frequency and strength of significant co-movement. Fourth, we model this bounded, zero-inflated dependent variable using Bayesian zero-inflated beta (ZIB) regression with a random-effects within--between (REWB) decomposition \cite{mundlak1978pooling,chamberlain1982multivariate}, separating within-dyad dynamics from between-dyad differences.

This approach allows us to test whether the correlates of synchronization differ across frequency bands, specifically, whether institutional integration (EU/EMU membership) and WB effects shows different associations at the different frequencies.

We begin by explaining how wavelet coherence measures co-movement in the time--frequency domain.

\subsection{Wavelet Coherence: Measuring Co-movement Across Frequencies}

Wavelet analysis decomposes a time series into the time–frequency domain. In contrast to Fourier analysis, which assumes stationarity\footnote{It is assumed that the frequency content and statistical properties don't change over time and that the power spectrum fully characterizes the process.} and captures only global frequency content, wavelet analysis uses localized basis functions that can track how spectral properties change over time.

The core object of interest is wavelet coherence, denoted $R_{xy}(s,\tau)$, which measures the localized correlation between two time series $x(t)$ and $y(t)$ at scale $s$ (corresponding to a specific period or frequency) and time location $\tau$. Following \cite{soares2011business}, we implement this using the Morlet wavelet, a complex-valued function that provides both amplitude (strength of co-movement) and phase (lead--lag relationships). Formally, coherence is expressed as:
\begin{equation}
R_{xy}(s,\tau)
\;=\;
\frac{\big|\,S\!\big(W_{xy}(s,\tau)\big)\big|}
{\big(S(|W_x(s,\tau)|^2)\big)^{1/2}\,
 \big(S(|W_y(s,\tau)|^2)\big)^{1/2}},
\label{eq:coherency}
\end{equation}
where $W_{xy}$ is the cross-wavelet transform (analogous to a localized cross-spectrum), $S(\cdot)$ denotes smoothing in both time and scale to ensure statistical robustness, and the denominator normalizes by the wavelet power of each series individually. The result, $R_{xy}(s,\tau)$, ranges from 0 to 1 and is analogous to a localized squared correlation measure in the time-frequency domain. In other words, this measures how strongly two countries' industrial production cycles move together at a specific frequency band and time window. A high coherence at the 3-year scale in 2008 would indicate that the countries experienced synchronized fluctuations during the financial crisis at that specific frequency.

Mathematical details for the continuous wavelet transform, wavelet power spectrum, cross-wavelets, and phase differences can be found in Appendix~\ref{app:wavelet_math}.

\subsection{Constructing the Dependent Variable: Annual In-Phase Synchronization}\label{sec:sync_dv}

Wavelet coherence produces rich monthly time--frequency information for each country pair. However, we need to compress it into a tractable dependent variable. In particular, we aggregate monthly coherence within two frequency bands, short-run (1.5--4.5 years) capturing common business-cycle fluctuations, and long-run (4.5--8.5 years) capturing structural co-movement, to construct an annual in-phase synchronization index.

This aggregation is done as follows. First, for each month $\tau$ and dyad $(a,b)$, we average coherence $R_{ab}(s,\tau)$ across all scales $s$ within a band $\mathcal{B}_b$, weighting by the inverse period $1/P_s$ to prevent longer cycles from mechanically dominating:
\begin{equation}
C^{(b)}_{ab,\tau}
\;=\;
\sum_{s \in \mathcal{B}_b} w_s\,R_{ab}(s,\tau),
\qquad
w_s \propto \frac{1}{P_s}, \;\; \sum w_s = 1.
\end{equation}
Without this weighting, a 4.5-year cycle would contribute three times as much as a 1.5-year cycle simply because it spans more time, even if both have equal coherence. The weighting ensures a frequency-balanced average (see Appendix~\ref{app:data} for details).

Second, we retain only months where coherence is statistically significant ($p \le 0.05$) and inside the cone of influence (COI), defining a monthly indicator:
\begin{equation}
I^{(b)}_{ab,\tau}
\;=\;
\mathbf{1}\!\left(\exists\, s \in \mathcal{B}_b : p_{ab}(s,\tau) \le 0.05 \;\;\wedge\;\; (s,\tau)\ \text{inside COI}\right).
\end{equation}
This filtering removes noise (insignificant coherence) and edge effects (COI; unreliable wavelet coefficients near time-series boundaries). However, it also introduces structural zeros: some dyad-years have no months with significant in-band coherence, representing genuine desynchronization rather than missing data. We address this zero-inflation explicitly in the regression model described in Section~\ref{sec:econometric-model}.

Third, let $\mathcal{M}^{(b)}_{ab,t}$ denote the eligible months in year $t$ (those inside the COI for band $b$). The annual in-phase synchronization measure is:
\begin{equation}
\mathrm{sync}^{(b)}_{ab,t}
\;=\;
\frac{1}{|\mathcal{M}^{(b)}_{ab,t}|}
\sum_{\tau \in \mathcal{M}^{(b)}_{ab,t}} I^{(b)}_{ab,\tau}\,C^{(b)}_{ab,\tau}
\;=\;
\mathrm{share}^{(b)}_{ab,t}\cdot \overline{C}^{(b)}_{ab,t},
\label{eq:sync_inphase}
\end{equation}
where $\mathrm{share}^{(b)}_{ab,t}$ is the fraction of eligible months with significant coherence and $\overline{C}^{(b)}_{ab,t}$ is the mean coherence conditional on significance. In other words, annual synchronization rises when a larger share of months exhibit significant co-movement and when coherence is stronger in those months. Values near zero indicate rare or weak co-movement; values near one indicate sustained, strong synchronization.

This construction necessarily abstracts from within-year timing (which months were synchronized), phase direction (which country leads), and scale-specific patterns (whether synchronization occurs at 2-year vs 4-year cycles within the short band). Future work could retain more wavelet information by modeling lead--lag measures ($\Delta T$ from Appendix~\ref{app:wavelet_math}), signed coherence, or higher-frequency panels. We also drop dyad-years with fewer than 9 total months or fewer than 6 eligible months outside the COI, removing spurious zeros from low coverage while retaining structural zeros that reflect genuine desynchronization (see Appendix~\ref{app:data} for filtering details).

\subsection{Econometric model: Zero-Inflated Beta Regression}
\label{sec:econometric-model}

We now specify a regression model accommodating the distributional features of the dependent variable.

The synchronization measure poses three modeling challenges. First, synchronization is a proportion, confined to the unit interval. Linear regression (OLS) can predict values outside $\left[0,1 \right]$, while logit or probit models are designed for binary outcomes (0 or 1 only) and cannot handle continuous values in the open interval (0,1). The beta distribution naturally accommodates continuous proportions. Second, approximately 15--20\% of dyad-years have exactly zero synchronization, not because of rounding or measurement error, but because no months in that year exhibited statistically significant coherence. These are structural zeros representing genuine desynchronization. A standard beta regression cannot accommodate exact zeros (the beta distribution has support on the open interval (0,1)). Third, some dyad-years cluster tightly around their mean synchronization (low variance), while others are highly dispersed (high variance). Beta regression allows the precision parameter $\phi$ to vary with covariates, accommodating this heterogeneity.

The zero-inflated beta (ZIB) model \cite{OspinaFerrari2012ZOIB} addresses all three challenges. It specifies a two-component mixture: with probability $\pi_{gt}$, synchronization is exactly zero (the dyad experiences complete desynchronization); with probability $1-\pi_{gt}$, synchronization follows a beta distribution on (0,1) with mean $\mu_{gt}$ and precision $\phi_{gt}$. We estimate this model using Bayesian methods via the \texttt{brms} package \cite{Buerkner2017brms} with a \texttt{Stan} backend \cite{Carpenter2017Stan}. Our baseline prior structure is a regularizing specification used throughout the main European short-run and long-run tables, as well as the long-run robustness analysis. Specifically, we place $\mathcal{N}(0,0.5)$ priors on population-level coefficients in the mean, precision, and zero-inflation equations; a Student-$t(3,0,2.5)$ prior on the mean intercept; $\mathcal{N}(0,1)$ and $\mathcal{N}(0,1.5)$ priors on the precision and zero-inflation intercepts; an Exponential$(1)$ prior on random-effect standard deviations; and an LKJ$(2)$ prior on random-effect correlations. For the Western Balkans (WB) analyses reported in the main text, we use an intermediate prior structure that relaxes coefficient shrinkage to $\mathcal{N}(0,1)$ for the population-level terms while keeping the same intercept, standard-deviation, and correlation priors, because those specifications are estimated on smaller effective samples and include many interaction terms. The appendix then reports the WB tables under both the regularizing prior structure and the default \texttt{brms} prior structure, under which the main population-level coefficients in the mean, precision, and zero-inflation equations receive no regularizing prior and are therefore estimated without coefficient shrinkage; the mean and precision intercepts receive Student-$t(3,0,2.5)$ priors, the zero-inflation intercept receives a logistic$(0,1)$ prior, random-effect standard deviations receive positive Student-$t(3,0,2.5)$ priors, and random-effect correlations receive an LKJ$(1)$ prior, so that prior sensitivity can be assessed directly. Table~\ref{tab:prior_structures} summarizes these three prior regimes and their use across the paper.

\begin{table}[htbp]
\centering
\scriptsize
\caption{Summary of prior structures used in the analysis}
\label{tab:prior_structures}
\begin{tabular}{p{2.9cm}p{4.6cm}p{5.2cm}}
\toprule
Prior structure & Population-level coefficient prior & Used for \\
\midrule
Strong regularization & $\mathcal{N}(0,0.5)$ in the mean, precision, and zero-inflation equations & Main European short-run and long-run tables, plus long-run robustness tables \\
Moderate regularization & $\mathcal{N}(0,1)$ in the mean, precision, and zero-inflation equations & Main-text WB convergence and WB three-way tables \\
No coefficient regularization & No regularizing prior on population-level coefficients & Appendix WB prior-sensitivity tables only \\
\bottomrule
\end{tabular}
\end{table}

For each dyad $g=(a,b)$ and year $t$, let $y_{gt}$ denote the synchronization measure in \eqref{eq:sync_inphase}. Formally, the outcome follows:
\begin{equation}
y_{gt} \sim
\begin{cases}
0, & \text{with probability } \pi_{gt},\\
\mathrm{Beta}(\mu_{gt}, \phi_{gt}), & \text{with probability } 1-\pi_{gt},
\end{cases}
\end{equation}
with logit links for $\mu_{gt}$ (conditional mean given non-zero) and $\pi_{gt}$ (zero probability), and a log link for $\phi_{gt}$ (precision):
\begin{equation}
\text{logit}(\mu_{gt}) = \eta^{(\mu)}_{gt}, \qquad
\text{logit}(\pi_{gt}) = \eta^{(\pi)}_{gt}, \qquad
\log(\phi_{gt}) = \eta^{(\phi)}_{gt}.
\end{equation}
The model estimates three distinct processes: (1) the conditional mean of synchronization when positive, (2) the probability of complete desynchronization, and (3) the precision (inverse dispersion) of the beta distribution. Higher $\phi$ corresponds to tighter clustering around $\mu$.

To separate within-dyad dynamics from between-dyad differences, we decompose each time-varying covariate $x_{gt}$ into two components:
\begin{equation}
x^{w}_{gt} = x_{gt} - \bar{x}_{g}, \qquad
x^{b}_{gt} = \bar{x}_{g} - \bar{x},
\end{equation}
where $\bar{x}_{g} = T^{-1}\sum_t x_{gt}$ is the dyad-specific time average and $\bar{x}$ is the grand mean. This yields within-dyad deviations ($x^w$) and between-dyad deviations ($x^b$), which enter the linear predictors separately.

The within-dyad effect captures how changes over time within a given pair are associated with synchronization. For example, when Germany--France trade intensity rises 5 percentage points above their 20-year average, does their synchronization increase? This isolates time-series variation, holding constant all fixed dyad characteristics. The between-dyad effect captures how baseline differences across pairs are associated with synchronization. For example, are high-trade dyads (e.g., Germany--Netherlands with 15\% average trade intensity) structurally more synchronized than low-trade dyads (e.g., Portugal--Finland with 2\% average trade intensity)? This isolates cross-sectional variation. For gap variables (e.g., specialization distance, fiscal gaps), the within effect measures how a widening or narrowing gap relative to that dyad's usual gap relates to synchronization, while the between effect compares dyads that are structurally more or less different on average.

The linear predictors for the mean and zero-inflation components are:
\begin{equation}
\eta^{(\mu)}_{gt} = \alpha_{\mu} + X^{w}_{gt}\beta_{\mu}^{w} + X^{b}_{gt}\beta_{\mu}^{b}
                  + Z_{gt}\gamma_{\mu} + u_{g} + \delta_{t},
\end{equation}
\begin{equation}
\eta^{(\pi)}_{gt} = \alpha_{\pi} + X^{w}_{gt}\beta_{\pi}^{w} + X^{b}_{gt}\beta_{\pi}^{b}
                  + Z_{gt}\gamma_{\pi} + u_{g} + \delta_{t},
\end{equation}
where $X^w$ and $X^b$ are matrices of within and between components for time-varying covariates, $Z_{gt}$ includes the lagged dependent variable; EU/EMU indicators are REWB-decomposed and therefore enter through $X^w_{gt}$ and $X^b_{gt}$, $u_{g}$ is a dyad-specific random intercept capturing unobserved dyad heterogeneity, and $\delta_{t}$ are year fixed effects. The precision model $\eta^{(\phi)}_{gt}$ uses an intercept, year effects, and a dyad random intercept. For a detailed illustration of how the within and between coefficients ($\beta^{w}$ and $\beta^{b}$) are interpreted for both continuous (e.g., trade) and binary institutional variables (e.g., EU/EMU membership), see Appendix~\ref{app:illustrative-example}.

\section{Results}\label{sec:results}

We present our empirical analysis in four steps. We begin by describing the time--frequency patterns of business-cycle co-movement with the EU benchmark using wavelet coherence and phase dynamics, which motivates distinguishing short-run (1.5--4.5 years) from long-run (4.5--8.5 years) synchronization. We then quantify the correlates of synchronization in each band using panel ZOIB models with an REWB decomposition, first for the short-run band and then for the long-run band to highlight horizon-specific differences in the estimated associations. Finally, we turn to a focused Western Balkans convergence analysis that contrasts EU--WB with EU--EU dyads and allows for country-specific heterogeneity in the channels underlying convergence across frequency bands.

\subsection{Wavelet evidence of synchronization}

First, we analyze the time--frequency structure of business-cycle co-movement with an EU benchmark using wavelet coherence and phase dynamics, which, as will be shown, motivates our separation of short-run (1.5--4.5 years) and long-run (4.5--8.5 years) synchronization. We construct the EU benchmark as a GDP-weighted average across all EU members.

\begin{figure}[htbp]
    \centering
    \includegraphics[width=\textwidth]{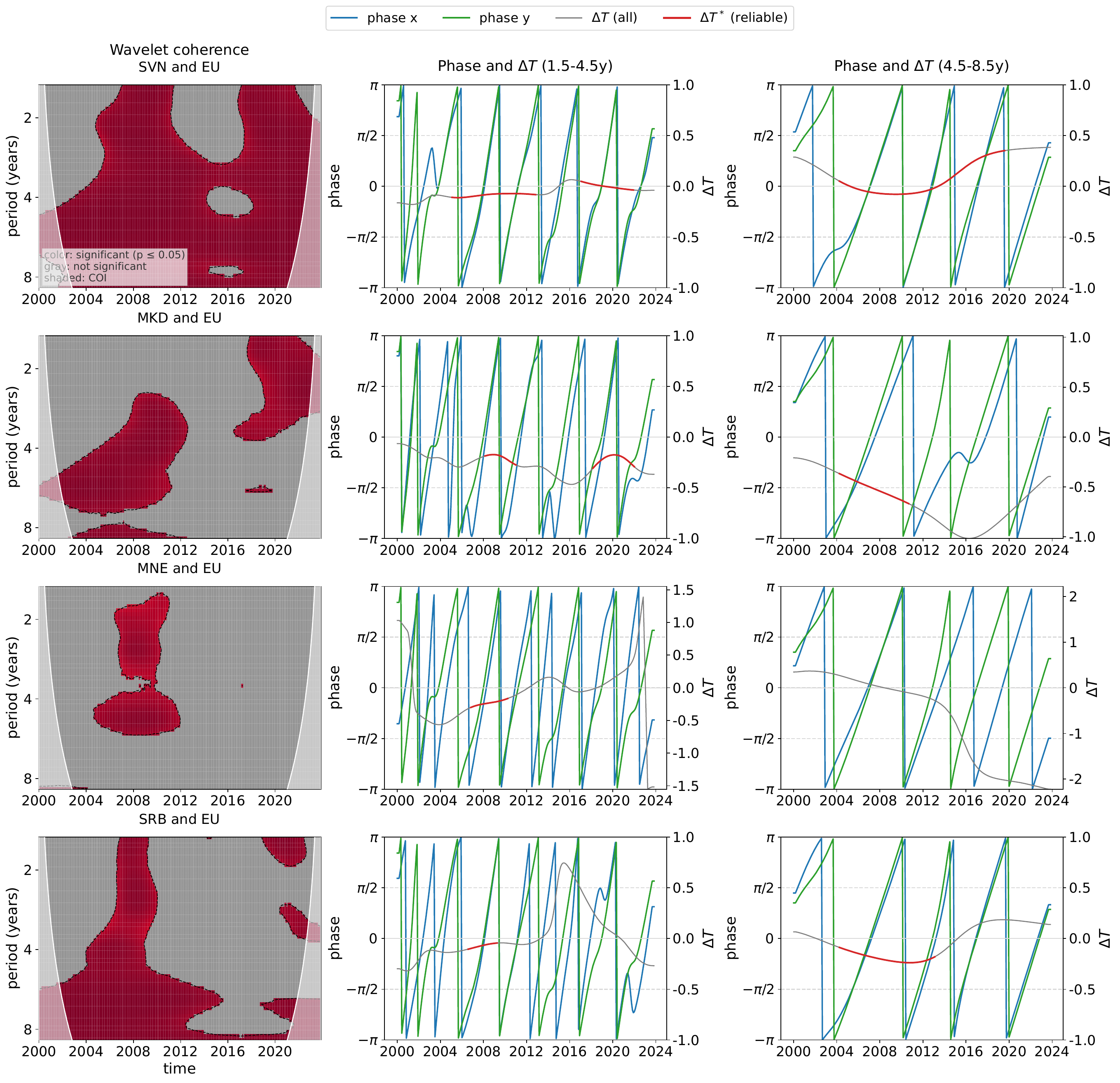}
    \caption{Wavelet coherence and phase dynamics relative to the EU aggregate. Each row corresponds to a country pair (SVN, MKD, MNE, SRB versus EU weighted average). Left column: coherence (color scale 0--1; warmer colors indicate higher coherence) with a dashed contour for $p \le 0.05$ and the cone of influence (COI) shaded in light gray; arrows encode phase differences. Middle and right columns: phase differences for the 1.5--4.5 and 4.5--8.5 year bands; the secondary axis overlays $\Delta T$ (gray) and reliable $\Delta T^*$ (red).}
    \label{fig:wavelet_grid}
\end{figure}

Figure~\ref{fig:wavelet_grid} summarizes wavelet coherence and phase dynamics between the EU benchmark and our three focal Western Balkan/neighboring economies (MKD, MNE, SRB). We also add Slovenia (SVN) to the example, to illustrate the differences between a Balkan EU member state and the candidates. 

The initial results in Fig.~\ref{fig:wavelet_grid} show that synchronization with the EU benchmark is generally stronger and more persistent at long-run frequencies (4.5--8.5 years) than at short-run frequencies (1.5--4.5 years). The result is heterogeneous across countries: Slovenia displays the strongest and most stable co-movement, Montenegro the weakest, Serbia shows clearer long-run synchronization, and N. Macedonia exhibits mixed results. Taken together, the figure motivates our focus on long-run versus short-run synchronization intensity in the panel regressions.

We also find that lead--lag dynamics are intermittent rather than fixed. That is, using reliable segments only ($\Delta T^*$), we observe that the pooled median absolute lead--lag is modest in both bands (0.10 years in the short band and 0.19 years in the longer band). Pair-specific medians are below 0.25 years in the short band and below 0.52 years in the longer band. These results indicate that synchronization strengthens at longer horizons, but directional leadership is episodic and not stable over time.

This suggests that the main cross-country signal is the intensity and prevalence of in-phase co-movement, not a persistent leader--follower structure. Therefore, in our regression analysis, we model annual synchronization levels (share $\times$ coherence) as the outcome of interest (see Eq.~\ref{eq:sync_inphase}).

\subsection{Determinants of short-run synchronization}

Next, we explore the determinants of short-run synchronization. We do this by performing a ZOIB regression analysis modeling the short band in-phase co-movement, as explained in Sections~\ref{sec:sync_dv} and~\ref{sec:econometric-model}.

Table~\ref{tab:zoib_results} reports the ZOIB estimates for the short-period band (1.5--4.5 years) with REWB decomposition. Columns~(1-3) present three baseline specifications that differ in lag structure: MAIN includes the lagged dependent variable in both the mean ($\mu$) and zero-inflation ($zi$) equations, SPEC A includes it only in $zi$, and SPEC B omits it entirely. Columns~(4)--(9) present robustness variants that retain the MAIN lag structure while removing or regrouping subsets of covariates. In particular, MAIN is the benchmark model, which includes the full covariate set and lagged synchronization in both the mean and zero-inflation equations; SPEC A and SPEC B then progressively relax this persistence structure by restricting the lag to the zero-inflation component and subsequently removing it altogether. The remaining specifications examine whether the baseline results survive under alternative choices: CORE retains only the central theoretical variables, providing a simplified specification that tests whether the main results emerge from the core theoretical channels alone; NO FIN excludes financial controls, allowing us to assess whether the baseline associations are driven by financial integration; NO TRADE removes trade intensity, testing whether EU/EMU effects are simply proxies for trade linkages; NO EU omits EU/EMU institutional indicators, thus showing what predicts synchronization in the absence of formal integration variables; POLICY focuses on more policy-relevant covariates, asking whether the results can be traced to variables more directly influenced by government decisions; and STRUCT isolates structural economic factors, assessing whether synchronization reflects underlying economic similarity rather than institutional membership. Taken together, these models show whether the estimated associations are sensitive to the treatment of persistence or instead reflect more robust links with institutional integration, trade, policy, and structural similarity.

\begin{table}[htbp]
\footnotesize
\centering
\renewcommand{\arraystretch}{0.82}
\caption{Short-run ZOIB Results: Determinants of synchronization (1.5-4.5 years)}
\label{tab:zoib_results}
\resizebox{\textwidth}{!}{%
\begin{tabular}{lccccccccc}
\toprule
 & MAIN & SPEC A & SPEC B & CORE & NO FIN & NO TRADE & NO EU & POLICY & STRUCT \\
\midrule
\multicolumn{10}{l}{\textit{Panel A: Mean Synchronization ($\mu$)}} \\
\addlinespace[0.2em]
\multicolumn{10}{l}{\textit{Within-dyad effects}} \\
\quad Lagged synchronization &  0.808*** & — & — &  0.816*** &  0.812*** &  0.809*** &  0.812*** &  0.817*** &  0.819*** \\
 & {\scriptsize [ 0.769,  0.847]} &  &  & {\scriptsize [ 0.779,  0.854]} & {\scriptsize [ 0.774,  0.851]} & {\scriptsize [ 0.771,  0.848]} & {\scriptsize [ 0.772,  0.851]} & {\scriptsize [ 0.779,  0.854]} & {\scriptsize [ 0.779,  0.859]} \\
\quad Both EU (non-EMU) &  0.253*** &  0.311*** &  0.312*** &  0.280*** &  0.250*** &  0.264*** & — &  0.268*** & — \\
 & {\scriptsize [ 0.130,  0.377]} & {\scriptsize [ 0.161,  0.452]} & {\scriptsize [ 0.171,  0.453]} & {\scriptsize [ 0.157,  0.402]} & {\scriptsize [ 0.122,  0.375]} & {\scriptsize [ 0.145,  0.388]} &  & {\scriptsize [ 0.145,  0.390]} &  \\
\quad Both EMU &  0.234** &  0.254** &  0.258** &  0.229*** &  0.223** &  0.236** & — &  0.224** & — \\
 & {\scriptsize [ 0.060,  0.408]} & {\scriptsize [ 0.053,  0.459]} & {\scriptsize [ 0.057,  0.453]} & {\scriptsize [ 0.055,  0.401]} & {\scriptsize [ 0.043,  0.399]} & {\scriptsize [ 0.066,  0.412]} &  & {\scriptsize [ 0.046,  0.397]} &  \\
\quad Trade Intensity &  0.043 &  0.199** &  0.197** &  0.010 &  0.022 & — &  0.073 & — &  0.059 \\
 & {\scriptsize [-0.110,  0.196]} & {\scriptsize [ 0.033,  0.371]} & {\scriptsize [ 0.022,  0.376]} & {\scriptsize [-0.138,  0.156]} & {\scriptsize [-0.124,  0.174]} &  & {\scriptsize [-0.079,  0.225]} &  & {\scriptsize [-0.095,  0.215]} \\
\quad Fiscal gap &  0.114*** &  0.171*** &  0.171*** & — &  0.104*** &  0.109*** &  0.110*** &  0.094*** & — \\
 & {\scriptsize [ 0.063,  0.168]} & {\scriptsize [ 0.111,  0.230]} & {\scriptsize [ 0.108,  0.231]} &  & {\scriptsize [ 0.053,  0.157]} & {\scriptsize [ 0.057,  0.161]} & {\scriptsize [ 0.056,  0.164]} & {\scriptsize [ 0.042,  0.145]} &  \\
\quad Specialization distance & -0.173*** & -0.187*** & -0.186*** & -0.138*** & -0.168*** & -0.173*** & -0.185*** & -0.156*** & -0.163*** \\
 & {\scriptsize [-0.246, -0.104]} & {\scriptsize [-0.265, -0.108]} & {\scriptsize [-0.268, -0.106]} & {\scriptsize [-0.207, -0.069]} & {\scriptsize [-0.237, -0.096]} & {\scriptsize [-0.245, -0.103]} & {\scriptsize [-0.254, -0.118]} & {\scriptsize [-0.223, -0.088]} & {\scriptsize [-0.232, -0.093]} \\
\quad Fin Open &  0.058 &  0.189*** &  0.190*** & — & — &  0.053 &  0.051 & — & — \\
 & {\scriptsize [-0.025,  0.142]} & {\scriptsize [ 0.082,  0.294]} & {\scriptsize [ 0.080,  0.301]} &  &  & {\scriptsize [-0.029,  0.134]} & {\scriptsize [-0.031,  0.136]} &  &  \\
\quad Urbanization diff. &  0.067 &  0.125 &  0.137 & — &  0.064 &  0.087 &  0.024 & — & -0.027 \\
 & {\scriptsize [-0.131,  0.264]} & {\scriptsize [-0.105,  0.359]} & {\scriptsize [-0.088,  0.364]} &  & {\scriptsize [-0.133,  0.261]} & {\scriptsize [-0.107,  0.278]} & {\scriptsize [-0.172,  0.223]} &  & {\scriptsize [-0.221,  0.170]} \\
\quad Remittances diff. &  0.092*** &  0.162*** &  0.164*** & — &  0.090** &  0.093** &  0.125*** & — & — \\
 & {\scriptsize [ 0.023,  0.163]} & {\scriptsize [ 0.085,  0.240]} & {\scriptsize [ 0.086,  0.241]} &  & {\scriptsize [ 0.020,  0.161]} & {\scriptsize [ 0.020,  0.166]} & {\scriptsize [ 0.056,  0.195]} &  &  \\
\quad Inflation diff. &  0.021 &  0.027 &  0.026 & — &  0.021 &  0.020 &  0.014 & — &  0.007 \\
 & {\scriptsize [-0.013,  0.055]} & {\scriptsize [-0.009,  0.064]} & {\scriptsize [-0.010,  0.062]} &  & {\scriptsize [-0.021,  0.062]} & {\scriptsize [-0.020,  0.061]} & {\scriptsize [-0.027,  0.056]} &  & {\scriptsize [-0.033,  0.047]} \\
\addlinespace[0.2em]
\multicolumn{10}{l}{\textit{Between-dyad effects}} \\
\quad Both EU (non-EMU) &  0.593*** &  0.560*** &  0.603*** &  0.425*** &  0.580*** &  0.573*** & — &  0.478*** & — \\
 & {\scriptsize [ 0.383,  0.806]} & {\scriptsize [ 0.354,  0.767]} & {\scriptsize [ 0.390,  0.819]} & {\scriptsize [ 0.249,  0.595]} & {\scriptsize [ 0.358,  0.789]} & {\scriptsize [ 0.356,  0.790]} &  & {\scriptsize [ 0.299,  0.658]} &  \\
\quad Both EMU &  0.555*** &  0.459*** &  0.510*** &  0.463*** &  0.548*** &  0.565*** & — &  0.500*** & — \\
 & {\scriptsize [ 0.353,  0.758]} & {\scriptsize [ 0.252,  0.662]} & {\scriptsize [ 0.293,  0.722]} & {\scriptsize [ 0.299,  0.628]} & {\scriptsize [ 0.341,  0.753]} & {\scriptsize [ 0.350,  0.773]} &  & {\scriptsize [ 0.336,  0.667]} &  \\
\quad Trade Intensity &  0.107*** &  0.102*** &  0.101*** &  0.101*** &  0.109*** & — &  0.120*** & — &  0.123*** \\
 & {\scriptsize [ 0.044,  0.170]} & {\scriptsize [ 0.043,  0.159]} & {\scriptsize [ 0.043,  0.159]} & {\scriptsize [ 0.042,  0.163]} & {\scriptsize [ 0.049,  0.170]} &  & {\scriptsize [ 0.057,  0.184]} &  & {\scriptsize [ 0.063,  0.185]} \\
\quad Fiscal gap & -0.057** & -0.050* & -0.047* & — & -0.048* & -0.064** & -0.056** & -0.049** & — \\
 & {\scriptsize [-0.107, -0.008]} & {\scriptsize [-0.105,  0.004]} & {\scriptsize [-0.102,  0.008]} &  & {\scriptsize [-0.099,  0.002]} & {\scriptsize [-0.116, -0.014]} & {\scriptsize [-0.107, -0.005]} & {\scriptsize [-0.098, -0.001]} &  \\
\quad Specialization distance & -0.011 & -0.025 & -0.024 & -0.022 & -0.011 & -0.039 & -0.056 & -0.042 & -0.030 \\
 & {\scriptsize [-0.101,  0.076]} & {\scriptsize [-0.110,  0.057]} & {\scriptsize [-0.109,  0.061]} & {\scriptsize [-0.088,  0.042]} & {\scriptsize [-0.084,  0.061]} & {\scriptsize [-0.123,  0.048]} & {\scriptsize [-0.148,  0.034]} & {\scriptsize [-0.108,  0.023]} & {\scriptsize [-0.106,  0.045]} \\
\quad Fin Open & -0.019 &  0.031 &  0.025 & — & — & -0.021 &  0.024 & — & — \\
 & {\scriptsize [-0.095,  0.057]} & {\scriptsize [-0.041,  0.102]} & {\scriptsize [-0.046,  0.098]} &  &  & {\scriptsize [-0.098,  0.054]} & {\scriptsize [-0.052,  0.103]} &  &  \\
\quad Urbanization diff. &  0.039* &  0.063*** &  0.070*** & — &  0.033 &  0.037 &  0.034 & — &  0.016 \\
 & {\scriptsize [-0.006,  0.085]} & {\scriptsize [ 0.016,  0.113]} & {\scriptsize [ 0.023,  0.118]} &  & {\scriptsize [-0.011,  0.076]} & {\scriptsize [-0.008,  0.081]} & {\scriptsize [-0.012,  0.080]} &  & {\scriptsize [-0.027,  0.061]} \\
\quad Remittances diff. &  0.091*** &  0.130*** &  0.131*** & — &  0.083** &  0.073** &  0.013 & — & — \\
 & {\scriptsize [ 0.026,  0.158]} & {\scriptsize [ 0.063,  0.200]} & {\scriptsize [ 0.059,  0.204]} &  & {\scriptsize [ 0.014,  0.152]} & {\scriptsize [ 0.008,  0.142]} & {\scriptsize [-0.038,  0.066]} &  &  \\
\quad Inflation diff. & -0.086* & -0.108** & -0.095* & — & -0.080* & -0.050 & -0.059 & — & -0.119** \\
 & {\scriptsize [-0.181,  0.011]} & {\scriptsize [-0.212, -0.007]} & {\scriptsize [-0.198,  0.006]} &  & {\scriptsize [-0.175,  0.014]} & {\scriptsize [-0.150,  0.050]} & {\scriptsize [-0.156,  0.037]} &  & {\scriptsize [-0.207, -0.031]} \\
\midrule
\multicolumn{10}{l}{\textit{Panel B: Zero-Inflation (zi) --- Negative = reduces P(zero)}} \\
\addlinespace[0.2em]
\multicolumn{10}{l}{\textit{Within-dyad effects}} \\
\quad Lagged synchronization & -2.541*** & -2.540*** & — & -2.539*** & -2.543*** & -2.543*** & -2.553*** & -2.544*** & -2.554*** \\
 & {\scriptsize [-2.687, -2.394]} & {\scriptsize [-2.687, -2.397]} &  & {\scriptsize [-2.694, -2.387]} & {\scriptsize [-2.696, -2.394]} & {\scriptsize [-2.691, -2.399]} & {\scriptsize [-2.697, -2.409]} & {\scriptsize [-2.696, -2.398]} & {\scriptsize [-2.703, -2.412]} \\
\quad Both EU (non-EMU) & -0.375** & -0.358** & -0.470*** & -0.370** & -0.364** & -0.439*** & — & -0.432*** & — \\
 & {\scriptsize [-0.711, -0.047]} & {\scriptsize [-0.687, -0.033]} & {\scriptsize [-0.750, -0.193]} & {\scriptsize [-0.704, -0.047]} & {\scriptsize [-0.701, -0.033]} & {\scriptsize [-0.769, -0.112]} &  & {\scriptsize [-0.755, -0.105]} &  \\
\quad Both EMU & -0.322 & -0.290 & -0.280 & -0.317 & -0.312 & -0.375 & — & -0.360 & — \\
 & {\scriptsize [-0.823,  0.173]} & {\scriptsize [-0.788,  0.211]} & {\scriptsize [-0.701,  0.148]} & {\scriptsize [-0.807,  0.177]} & {\scriptsize [-0.791,  0.172]} & {\scriptsize [-0.859,  0.104]} &  & {\scriptsize [-0.833,  0.119]} &  \\
\quad Trade Intensity & -0.560** & -0.580** & -1.007*** & -0.576** & -0.570** & — & -0.677*** & — & -0.677*** \\
 & {\scriptsize [-1.032, -0.098]} & {\scriptsize [-1.042, -0.093]} & {\scriptsize [-1.382, -0.622]} & {\scriptsize [-1.036, -0.120]} & {\scriptsize [-1.040, -0.102]} &  & {\scriptsize [-1.137, -0.217]} &  & {\scriptsize [-1.124, -0.234]} \\
\quad Specialization distance &  0.240** &  0.251*** &  0.279*** &  0.243*** &  0.248*** &  0.254*** &  0.248*** &  0.249** &  0.249*** \\
 & {\scriptsize [ 0.053,  0.425]} & {\scriptsize [ 0.058,  0.439]} & {\scriptsize [ 0.132,  0.429]} & {\scriptsize [ 0.058,  0.428]} & {\scriptsize [ 0.059,  0.432]} & {\scriptsize [ 0.065,  0.440]} & {\scriptsize [ 0.066,  0.436]} & {\scriptsize [ 0.062,  0.434]} & {\scriptsize [ 0.060,  0.432]} \\
\quad Fin Open &  0.037 &  0.045 & -0.218* & — & — &  0.056 &  0.024 & — & — \\
 & {\scriptsize [-0.241,  0.314]} & {\scriptsize [-0.246,  0.334]} & {\scriptsize [-0.448,  0.008]} &  &  & {\scriptsize [-0.220,  0.328]} & {\scriptsize [-0.254,  0.304]} &  &  \\
\addlinespace[0.2em]
\multicolumn{10}{l}{\textit{Between-dyad effects}} \\
\quad Both EU (non-EMU) & -1.006*** & -1.052*** & -1.046*** & -0.984*** & -0.950*** & -1.029*** & — & -1.029*** & — \\
 & {\scriptsize [-1.458, -0.552]} & {\scriptsize [-1.512, -0.597]} & {\scriptsize [-1.412, -0.677]} & {\scriptsize [-1.427, -0.530]} & {\scriptsize [-1.405, -0.485]} & {\scriptsize [-1.501, -0.564]} &  & {\scriptsize [-1.498, -0.562]} &  \\
\quad Both EMU & -1.146*** & -1.160*** & -1.131*** & -0.987*** & -0.970*** & -1.220*** & — & -1.053*** & — \\
 & {\scriptsize [-1.599, -0.707]} & {\scriptsize [-1.629, -0.716]} & {\scriptsize [-1.481, -0.780]} & {\scriptsize [-1.441, -0.562]} & {\scriptsize [-1.407, -0.531]} & {\scriptsize [-1.662, -0.781]} &  & {\scriptsize [-1.494, -0.630]} &  \\
\quad Trade Intensity & -0.261*** & -0.279*** & -0.176*** & -0.267*** & -0.265*** & — & -0.298*** & — & -0.300*** \\
 & {\scriptsize [-0.439, -0.085]} & {\scriptsize [-0.459, -0.095]} & {\scriptsize [-0.307, -0.041]} & {\scriptsize [-0.443, -0.088]} & {\scriptsize [-0.440, -0.086]} &  & {\scriptsize [-0.479, -0.117]} &  & {\scriptsize [-0.477, -0.130]} \\
\quad Specialization distance & -0.122 & -0.131 & -0.075 &  0.024 &  0.019 & -0.084 &  0.001 &  0.088 &  0.091 \\
 & {\scriptsize [-0.350,  0.107]} & {\scriptsize [-0.358,  0.098]} & {\scriptsize [-0.250,  0.093]} & {\scriptsize [-0.170,  0.219]} & {\scriptsize [-0.183,  0.211]} & {\scriptsize [-0.302,  0.136]} & {\scriptsize [-0.227,  0.231]} & {\scriptsize [-0.101,  0.282]} & {\scriptsize [-0.106,  0.293]} \\
\quad Fin Open &  0.283*** &  0.261** &  0.221*** & — & — &  0.305*** &  0.168 & — & — \\
 & {\scriptsize [ 0.074,  0.489]} & {\scriptsize [ 0.039,  0.469]} & {\scriptsize [ 0.064,  0.374]} &  &  & {\scriptsize [ 0.099,  0.519]} & {\scriptsize [-0.043,  0.369]} &  &  \\
\midrule
\multicolumn{10}{l}{\textit{Panel C: Model Fit}} \\
ELPD-LOO & 1970.3 & 1308.8 & 320.8 & 1943.3 & 1959.5 & 1963.6 & 1950.4 & 1956.2 & 1941.8 \\
$\Delta$ ELPD (SE) & --- & -661.5 (43) & -1649.5 (56.5) & -27.0 (8.3) & -10.8 (4.5) & -6.7 (4.4) & -19.9 (7.6) & -14.1 (6.9) & -28.5 (10.2) \\
Observations & 7,067 & 7,067 & 7,067 & 7,067 & 7,067 & 7,067 & 7,067 & 7,067 & 7,067 \\
Dyads & 406 & 406 & 406 & 406 & 406 & 406 & 406 & 406 & 406 \\
\bottomrule
\end{tabular}
}
\begin{tablenotes}[flushleft]
\footnotesize
\item \textit{Notes}: Posterior means with 95\% credible intervals in brackets.
Stars indicate posterior credible-interval exclusion of zero: * 90\%, ** 95\%, *** 99\%. They are not p-values.
ELPD-LOO is the leave-one-out expected log predictive density, so higher values indicate better out-of-sample predictive fit. $\Delta$ ELPD is computed relative to MAIN; negative values indicate worse predictive fit than MAIN, and the value in parentheses is the standard error of that difference.
Within effects: deviations from dyad mean. Between effects: dyad means.
Year fixed effects and dyad random effects included.
\end{tablenotes}
\end{table}

We observe that short-run synchronization is highly persistent. The lagged dependent variable enters the mean equation with a posterior mean of value around $0.8$ where it is included, and the zero-inflation equation at values around $-2.5$, indicating that dyads with above-average coherence in one period tend to remain synchronized in the next, and that previously synchronized dyads are less likely to fall into the zero-coherence regime. These estimates are stable across all robustness specifications.

In terms of institutional integration, shared EU membership outside the EMU and EMU membership are associated with higher synchronization\footnote{Unless otherwise noted, the coefficient magnitudes discussed here refer to the MAIN specification in Table \ref{tab:zoib_results}; alternative specifications are mentioned only where they change the interpretation.}. Pairs in which both countries share EU membership outside the EMU show a mean coefficient of roughly $0.59$, and joint EMU membership is associated with a coefficient of comparable magnitude. The corresponding zero-inflation coefficients are strongly negative ($-1.01$ for EU membership outside the EMU, $-1.15$ for EMU), indicating that integrated dyads are much less likely to exhibit years of no synchronization. Within-dyad associations that capture changes in synchronization as dyads transition over time into joint membership are smaller but point in the same direction: positive in $\mu$ ($0.25$ for EU membership outside the EMU, $0.23$ for EMU) and generally negative in $zi$. Across the specifications that retain the institutional indicators, the between-dyad non-EMU EU and EMU coefficients remain large and statistically credible, while the within-dyad EMU coefficient in $\mu$ remains smaller than the between-dyad EMU coefficient but stays credibly positive throughout. In specifications where EU/EMU indicators are omitted by construction, namely NO EU and STRUCT, dropping them does not materially alter the main trade, fiscal, and specialization results. This suggests the EU/EMU variables are not absorbing most of the variation that would otherwise be attributed to the other regressors. We note that these associations do not establish causality; unobserved factors correlated with both EU membership and synchronization could account for part or all of these observations.

Trade intensity in the short run is primarily a between-dyad phenomenon. Country pairs with higher sample-period mean bilateral trade intensity than other pairs tend to have higher mean synchronization, whereas year-to-year deviations from a pair's own average trade intensity (the within-dyad component) are not robustly related to mean synchronization once persistence is controlled. However, higher trade intensity both within dyads over time and across dyads on average is associated with a lower probability of zero synchronization in the $zi$ equation. When the lagged dependent variable is removed (SPEC A and SPEC B), within-dyad trade becomes positively associated with $\mu$ as well, suggesting that its conditional association with mean synchronization is largely mediated through persistence. More generally, the trade coefficient is positive in the between-dyad $\mu$ equation across all specifications in which trade is included, and negative in the zero-inflation equation throughout. Specialization distance is consistently negative in $\mu$ and positive in $zi$ at the within-dyad level across all specifications in which it is included. This means that as dyads become more structurally different over time (within-dyad), their synchronization declines and their probability of complete de-synchronization rises, whereas the corresponding between-dyad specialization coefficients are much weaker and less stable.

Among the remaining variables, fiscal gaps show mixed results in the specifications that retain this covariate. The within-dyad association with the mean is positive, while the between-dyad association is negative. This implies that when a given dyad's fiscal gap is above its own sample-period average, synchronization is slightly higher, but dyads that are persistently more fiscally dissimilar than other dyads tend to be less synchronized on average. One interpretation is that short-run fiscal movements within pairs may co-move with common cyclical shocks, whereas persistent cross-dyad fiscal difference may imply deeper institutional or structural differences. Remittance gaps show a modest positive association with synchronization in most specifications where they are included, and inflation gaps tend to be negatively associated with between-dyad synchronization, most clearly in STRUCT but not uniformly across specifications, although neither is robust across all specifications. Overall, the general similarity between MAIN and the alternative specifications indicates that the short-run results are reasonably robust to changes in the control set.

\subsection{Determinants of long-run synchronization}

Next, we investigate the long-run synchronization determinants. 
Table~\ref{tab:zoib_results_long} reports the ZOIB estimates for the 4.5--8.5 years band. As in the short-run band, synchronization is strongly persistent: in MAIN, the lagged dependent variable enters the mean equation at approximately $0.76$ and the zero-inflation equation at $-2.94$.

The results are different from the short-run results in several aspects. First, within-dyad trade intensity is positively associated with the conditional mean ($0.39$), suggesting that sustained bilateral trade deepening is a relevant correlate of synchronization in the longer frequency-band. Second, the within-dyad non-EMU EU and EMU coefficients reverse sign: once persistence is controlled for, transitions over time into joint EU/EMU membership are associated with lower conditional mean synchronization ($-0.16$ for non-EMU EU, $-0.26$ for EMU). Between-dyad non-EMU EU and EMU coefficients, which are large and precise in the short-run band, are indistinguishable from zero at this horizon. Third, structural divergence takes on greater importance at longer frequencies: the between-dyad specialization distance coefficient is $-0.14$ in the mean equation and $0.59$ in the zero-inflation equation, indicating that dyads with persistently different economic structures exhibit both lower synchronization and a substantially higher probability of zero synchronization. Finally, financial openness within dyads is negatively associated with $\mu$ ($-0.16$) a result broadly consistent with evidence for the Western Balkans that some cross-border financial channels, including FDI, bank flows, and remittances, may weaken business-cycle synchronization rather than strengthen it \cite{hildebrandt2015business}. Because our measure captures overall external openness rather than bilateral financial flows, the comparison should be interpreted as suggestive rather than one-to-one.

\begin{table}[htbp]
\centering
\caption{Long-run ZOIB Results: Determinants of synchronization (4.5--8.5 years)}
\label{tab:zoib_results_long}
\resizebox{\textwidth}{!}{%
\begin{tabular}{lcccccc}
\toprule
 & \multicolumn{3}{c}{Mean Synchronization ($\mu$)} & \multicolumn{3}{c}{Zero-Inflation ($zi$)} \\
\cmidrule(lr){2-4} \cmidrule(lr){5-7}
 & MAIN & SPEC A & SPEC B & MAIN & SPEC A & SPEC B \\
\midrule
\multicolumn{7}{l}{\textit{Within-dyad effects}} \\
\quad Lagged synchronization &  0.755*** & --- & --- & $-$2.940*** & $-$2.942*** & --- \\
 & [0.706, 0.803] &  &  & [-3.115, -2.770] & [-3.119, -2.766] &  \\
\quad Both EU (non-EMU) & $-$0.160** & $-$0.087 & $-$0.077 & $-$0.648*** & $-$0.651*** & $-$0.858*** \\
 & [-0.303, -0.019] & [-0.248, 0.068] & [-0.235, 0.082] & [-1.054, -0.252] & [-1.040, -0.249] & [-1.179, -0.539] \\
\quad Both EMU & $-$0.260*** & $-$0.144 & $-$0.125 & $-$0.155 & $-$0.124 & $-$0.640*** \\
 & [-0.454, -0.064] & [-0.364, 0.068] & [-0.349, 0.101] & [-0.709, 0.399] & [-0.674, 0.441] & [-1.098, -0.187] \\
\quad Trade Intensity &  0.386*** &  0.548*** &  0.542*** & $-$0.477* & $-$0.473* & $-$1.002*** \\
 & [0.208, 0.567] & [0.366, 0.732] & [0.355, 0.733] & [-1.022, 0.075] & [-1.014, 0.055] & [-1.420, -0.586] \\
\quad Fiscal gap & $-$0.055* & $-$0.078** & $-$0.079*** & --- & --- & --- \\
 & [-0.110, 0.003] & [-0.142, -0.015] & [-0.143, -0.017] &  &  &  \\
\quad Specialization distance &  0.030 &  0.019 &  0.024 &  0.519*** &  0.513*** &  0.611*** \\
 & [-0.046, 0.104] & [-0.064, 0.102] & [-0.058, 0.109] & [0.248, 0.800] & [0.242, 0.797] & [0.417, 0.808] \\
\quad Fin Open & $-$0.157*** & $-$0.118* & $-$0.114* &  0.156 &  0.161 & $-$0.090 \\
 & [-0.260, -0.053] & [-0.236, 0.002] & [-0.232, 0.005] & [-0.208, 0.515] & [-0.201, 0.539] & [-0.379, 0.197] \\
\quad Urbanization diff. &  0.187 &  0.462*** &  0.470*** & --- & --- & --- \\
 & [-0.052, 0.415] & [0.199, 0.721] & [0.211, 0.732] &  &  &  \\
\quad Remittances diff. & $-$0.055 & $-$0.029 & $-$0.031 & --- & --- & --- \\
 & [-0.138, 0.029] & [-0.121, 0.062] & [-0.122, 0.063] &  &  &  \\
\quad Inflation diff. &  0.032* &  0.049*** &  0.049*** & --- & --- & --- \\
 & [-0.001, 0.065] & [0.015, 0.084] & [0.015, 0.084] &  &  &  \\
\addlinespace[0.5em]
\multicolumn{7}{l}{\textit{Between-dyad effects}} \\
\quad Both EU (non-EMU) & $-$0.010 & $-$0.049 & $-$0.070 & $-$0.331 & $-$0.269 & $-$0.079 \\
 & [-0.268, 0.251] & [-0.301, 0.202] & [-0.321, 0.174] & [-0.957, 0.300] & [-0.915, 0.367] & [-0.589, 0.422] \\
\quad Both EMU & $-$0.051 & $-$0.028 & $-$0.051 & $-$0.108 & $-$0.139 &  0.065 \\
 & [-0.320, 0.221] & [-0.290, 0.234] & [-0.306, 0.208] & [-0.733, 0.544] & [-0.780, 0.497] & [-0.482, 0.610] \\
\quad Trade Intensity &  0.038 &  0.022 &  0.018 & $-$0.100 & $-$0.088 & $-$0.047 \\
 & [-0.045, 0.125] & [-0.052, 0.100] & [-0.055, 0.091] & [-0.400, 0.187] & [-0.384, 0.210] & [-0.263, 0.164] \\
\quad Fiscal gap &  0.019 &  0.024 &  0.021 & --- & --- & --- \\
 & [-0.049, 0.087] & [-0.045, 0.096] & [-0.046, 0.091] &  &  &  \\
\quad Specialization distance & $-$0.143** & $-$0.092* & $-$0.098* &  0.589*** &  0.582*** &  0.459*** \\
 & [-0.255, -0.032] & [-0.189, 0.007] & [-0.198, 0.001] & [0.216, 0.964] & [0.218, 0.947] & [0.180, 0.741] \\
\quad Fin Open &  0.066 &  0.056 &  0.058 & $-$0.150 & $-$0.127 & $-$0.146 \\
 & [-0.033, 0.170] & [-0.031, 0.144] & [-0.029, 0.148] & [-0.492, 0.195] & [-0.481, 0.213] & [-0.412, 0.119] \\
\quad Urbanization diff. & $-$0.019 & $-$0.017 & $-$0.015 & --- & --- & --- \\
 & [-0.080, 0.043] & [-0.082, 0.049] & [-0.079, 0.048] &  &  &  \\
\quad Remittances diff. & $-$0.023 & $-$0.002 & $-$0.030 & --- & --- & --- \\
 & [-0.107, 0.064] & [-0.091, 0.088] & [-0.116, 0.058] &  &  &  \\
\quad Inflation diff. &  0.100 &  0.016 &  0.036 & --- & --- & --- \\
 & [-0.045, 0.239] & [-0.138, 0.168] & [-0.112, 0.185] &  &  &  \\
\midrule
\multicolumn{7}{c}{\textit{Model Fit (whole model)}} \\
 & MAIN & SPEC A & SPEC B & & & \\
ELPD-LOO & 2555.4 & 2202.0 & 1109.7 & Obs & \multicolumn{2}{c}{6,507} \\
$\Delta$ ELPD (SE) & --- & -353.4 (41.5) & -1445.7 (56.2) & Dyads & \multicolumn{2}{c}{406} \\
\bottomrule
\end{tabular}
}
\begin{tablenotes}[flushleft]
\footnotesize
\item \textit{Notes}: Posterior means with 95\% credible intervals in brackets. $zi$: negative = reduces P(zero).
Stars indicate posterior credible-interval exclusion of zero: * 90\%, ** 95\%, *** 99\%. They are not p-values.
ELPD-LOO is the leave-one-out expected log predictive density, so higher values indicate better out-of-sample predictive fit. $\Delta$ ELPD is computed relative to MAIN; negative values indicate worse predictive fit than MAIN, and the value in parentheses is the standard error of that difference.
Within effects: deviations from dyad mean. Between effects: dyad means.
Year fixed effects and dyad random effects included.
\end{tablenotes}
\end{table}

The comparison between the results for the two frequency bands are summarized in Table~\ref{tab:short_long_comparison}. Institutional integration is strongly and positively associated with synchronization in the shorter band but reverses sign in the longer band in the within-dyad estimation, while between-dyad effects are negligible. Trade operates through different channels at each horizon: the between-dyad level of bilateral trade matters for short-run synchronization, while within-dyad trade deepening is relevant in the longer term. Specialization distance is associated with lower synchronization at both horizons, but the source of this association differs across frequencies: in the short run it comes mainly from within-dyad changes over time, whereas in the long run it comes through both a strong within-dyad zero-inflation margin and between-dyad differences in average specialization distance. Financial openness, which shows no clear short-run association, is negatively correlated with synchronization in the 4.5-8.5 years band.

\begin{table}[htbp]
\centering
\caption{Comparison of short-run vs.\ long-run determinants (MAIN specification)}
\label{tab:short_long_comparison}
\small
\begin{threeparttable}
\begin{tabular}{lccp{5cm}}
\toprule
Variable & Short & Long & Interpretation \\
\midrule
\multicolumn{4}{l}{\textit{Mean equation ($\mu$)}} \\
Lag & 0.81*** & 0.75*** & Less persistent in the long run \\
Trade (w) & n.s. & 0.39*** & Trade deepening over time matters mainly long-run \\
Trade (b) & 0.11*** & n.s. & Average trade differences across pairs matter only short-run \\
Spec.\ dist.\ (w) & $-$0.17*** & n.s. & Structural divergence over time lowers sync only short-run \\
Spec.\ dist.\ (b) & n.s. & $-$0.14** & Average structural similarity matters mainly long-run \\
Fin.\ open (w) & n.s. & $-$0.16*** & Financial opening over time lowers long-run sync \\
EU (w) & 0.25*** & $-$0.16** & EU effect over time flips sign \\
EMU (w) & 0.23** & $-$0.26*** & EMU effect over time flips sign \\
EU (b) & 0.59*** & n.s. & Average EU effect across pairs is short-run only \\
EMU (b) & 0.55*** & n.s. & Average EMU effect across pairs is short-run only \\
\midrule
\multicolumn{4}{l}{\textit{Zero-inflation ($zi$) --- negative = reduces P(zero)}} \\
Lag & $-$2.54*** & $-$2.94*** & Past sync lowers zero risk more strongly long-run \\
Spec.\ dist.\ (w) & 0.24** & 0.52*** & Structural divergence over time raises zero risk more strongly long-run \\
Spec.\ dist.\ (b) & n.s. & 0.59*** & Average structural dissimilarity raises zero risk only long-run \\
Trade (b) & $-$0.26*** & n.s. & Average trade across pairs lowers zero risk only short-run \\
EU (b) & $-$1.01*** & n.s. & Average EU status lowers zero risk only short-run \\
EMU (b) & $-$1.15*** & n.s. & Average EMU status lowers zero risk only short-run \\
\bottomrule
\end{tabular}
\begin{tablenotes}
\footnotesize
\item Notes: Row labels distinguish changes over time within the same dyad from differences in average values across dyads. Stars indicate posterior credible-interval exclusion of zero: * 90\%, ** 95\%, *** 99\%. They are not p-values.
\item Effective estimation sample: short-run N = 7,067, dyads = 406; long-run N = 6,507, dyads = 406.
\end{tablenotes}
\end{threeparttable}
\end{table}

Table~\ref{tab:robustness_long} examines the robustness of negative within-dyad EU/EMU associations at longer-term frequencies. Removing EU and EMU dummies entirely does not alter the other coefficients qualitatively (the lag remains at $0.75$, within-dyad trade at $0.35$, and between-dyad specialization distance remains negative at roughly $-0.12$ to $-0.15$). This indicates that the long-run sign-reversal pattern is not driven by collinearity or absorption by other regressors. The final (STRUCT) specification, which retains only structural variables, yields a similar result: specialization distance remains the strongest between-dyad correlate of longer-term synchronization. Across all three specifications, trade deepening and lower cross-dyad specialization distance emerge as more robust correlates of the longer-term synchronization than EU or EMU membership.

\begin{table}[htbp]
\centering
\caption{Long-run robustness: Key coefficients across specifications}
\label{tab:robustness_long}
\begin{threeparttable}
\begin{tabular}{lccc}
\toprule
& MAIN & NO EU & STRUCT \\
\midrule
\multicolumn{4}{l}{\textit{Mean equation ($\mu$)}} \\[3pt]
Lag (w) & 0.75*** & 0.75*** & 0.75*** \\
& [0.71, 0.80] & [0.70, 0.80] & [0.70, 0.80] \\[3pt]
Trade intensity (w) & 0.39*** & 0.35*** & 0.35*** \\
& [0.21, 0.57] & [0.17, 0.52] & [0.17, 0.52] \\[3pt]
Spec.\ distance (b) & $-$0.14** & $-$0.15*** & $-$0.12** \\
& [-0.25, -0.03] & [-0.26, -0.04] & [-0.21, -0.02] \\[3pt]
Fin.\ openness (w) & $-$0.16*** & $-$0.16*** & --- \\
& [-0.26, -0.05] & [-0.27, -0.05] \\[3pt]
Both EU non-EMU (w) & $-$0.16** & --- & --- \\
& [-0.30, -0.02] \\[3pt]
Both EMU (w) & $-$0.26*** & --- & --- \\
& [-0.45, -0.06] \\
\midrule
\multicolumn{4}{l}{\textit{Zero-inflation equation ($zi$)}} \\[3pt]
Lag (w) & $-$2.94*** & $-$2.94*** & $-$2.93*** \\
& [-3.11, -2.77] & [-3.11, -2.76] & [-3.11, -2.76] \\[3pt]
Trade intensity (w) & $-$0.48* & $-$0.60** & $-$0.61** \\
& [-1.02, 0.07] & [-1.14, -0.08] & [-1.13, -0.08] \\[3pt]
Spec.\ distance (b) & 0.59*** & 0.61*** & 0.55*** \\
& [0.22, 0.96] & [0.25, 0.97] & [0.21, 0.88] \\
\midrule
\multicolumn{4}{l}{\textit{Sample}} \\[3pt]
Observations & 6,507 & 6,507 & 6,507 \\
Dyads & 406 & 406 & 406 \\
\bottomrule
\end{tabular}
\begin{tablenotes}
\small
\item Notes: Posterior means with 95\% credible intervals in brackets. MAIN includes all covariates; NO EU removes EU/EMU dummies; STRUCT retains only structural variables (trade, specialization, urbanization, inflation). (w) = within-dyad; (b) = between-dyad. Stars indicate posterior credible-interval exclusion of zero: * 90\%, ** 95\%, *** 99\%. They are not p-values. Sample counts are the effective estimation samples used in each fitted column.
\end{tablenotes}
\end{threeparttable}
\end{table}

\subsection{Western Balkans convergence analysis}

Finally, we use our approach to examine whether Western Balkan (WB) economies converge toward EU benchmark levels of business cycle synchronization, and identify the channels through which this convergence operates across frequency bands. We address this by comparing EU--WB dyads (one EU member paired with MKD, SRB, or MNE) with EU--EU dyads, while excluding WB--WB pairs. For the WB analyses reported in the main text, we use the intermediate-prior specification; Appendix~\ref{app:wb_prior_sensitivity} reports the corresponding WB tables under both the regularizing and default-prior structures.

Table~\ref{tab:wb_convergence} reports the main EU--WB convergence estimates for short-run (1.5--4.5 years) and long-run (4.5--8.5 years) synchronization. At short-run frequencies, the EU--WB indicator is clearly negative in the mean equation ($-0.59$) and positive in the zero-inflation equation ($1.62$). This means that EU--WB pairs are associated with lower conditional synchronization and higher log-odds of falling into the zero-synchronization regime. The within-dyad trade interaction is positive in the mean equation ($1.38$) but is no longer credibly different from zero in the zero-inflation equation, indicating that periods of trade deepening within EU--WB dyads are associated with a more positive relationship with synchronization than in EU--EU dyads in the conditional mean. By contrast, between-dyad trade interactions are not credibly different from zero, and specialization-distance interactions in the mean equation remain close to zero in this band.

The long-run results differ in two ways. First, the EU--WB main gap narrows substantially: the mean-equation coefficient is $0.10$ and the zero-inflation coefficient is $0.60$; in both cases, the interval includes zero, so we do not find clear evidence of an average EU--WB difference relative to EU--EU dyads. Second, the interaction profile shifts away from trade and toward structural variables. Trade interactions are near zero and imprecise in both equations, whereas the within-dyad specialization-distance interaction in the mean equation is negative ($-0.26$). This sign implies that, relative to EU--EU dyads, reductions in sectoral dissimilarity within EU--WB dyads are associated with stronger long-run synchronization. Under the intermediate priors, the zero-inflation specialization interactions are not credibly different from zero, so the long-run structural evidence is concentrated in the conditional-mean equation rather than in the probability of falling into the zero-synchronization regime.

\begin{table}[htbp]
\centering
\caption{EU-WB convergence analysis: Short-run vs.\ long-run synchronization}
\label{tab:wb_convergence}
\begin{threeparttable}
\begin{tabular}{lcccc}
\toprule
& \multicolumn{2}{c}{Short-run (1.5--4.5y)} & \multicolumn{2}{c}{Long-run (4.5--8.5y)} \\
\cmidrule(lr){2-3} \cmidrule(lr){4-5}
Coefficient & Estimate & 95\% CI & Estimate & 95\% CI \\
\midrule
\multicolumn{5}{l}{\textit{Panel A: Mean equation ($\mu$)}} \\[3pt]
EU-WB main effect & $-$0.59*** & [$-$0.84, $-$0.33] & 0.10 & [$-$0.19, 0.40] \\
Trade intensity (w) $\times$ EU-WB & 1.38*** & [0.61, 2.15] & 0.02 & [$-$0.69, 0.72] \\
Trade intensity (b) $\times$ EU-WB & $-$0.03 & [$-$0.48, 0.44] & 0.01 & [$-$0.51, 0.51] \\
Spec.\ distance (w) $\times$ EU-WB & 0.01 & [$-$0.23, 0.25] & $-$0.26** & [$-$0.50, $-$0.02] \\
Spec.\ distance (b) $\times$ EU-WB & 0.06 & [$-$0.17, 0.30] & 0.00 & [$-$0.30, 0.29] \\
\midrule
\multicolumn{5}{l}{\textit{Panel B: Zero-inflation equation ($zi$)}} \\[3pt]
EU-WB main effect & 1.62*** & [1.00, 2.25] & 0.60 & [$-$0.33, 1.49] \\
Trade intensity (w) $\times$ EU-WB & $-$0.93 & [$-$2.53, 0.64] & 0.19 & [$-$1.59, 1.97] \\
Trade intensity (b) $\times$ EU-WB & 0.15 & [$-$1.03, 1.32] & 0.01 & [$-$1.44, 1.47] \\
Spec.\ distance (w) $\times$ EU-WB & $-$0.44 & [$-$1.00, 0.12] & $-$0.28 & [$-$1.07, 0.53] \\
Spec.\ distance (b) $\times$ EU-WB & $-$0.11 & [$-$0.71, 0.50] & $-$0.70 & [$-$1.65, 0.28] \\
\bottomrule
\end{tabular}
\begin{tablenotes}
\scriptsize
\item Notes: Estimates from the FULL INT specification with trade and specialization interactions. The table reports focal EU-WB coefficients only. Models are additionally estimated with controls: lagged dependent variable, main effects for trade intensity and specialization distance (within and between), fiscal-gap, urban-gap, remittance-gap, inflation-gap, and financial-openness terms in the mean equation, plus year fixed effects and dyad random intercepts; the zero-inflation equation includes lagged dependent variable, trade, specialization distance, financial openness, year fixed effects, and dyad random intercepts. EU-WB = 1 if exactly one country in the pair is Western Balkan (MKD, SRB, MNE); reference group is EU-EU pairs. WB-WB pairs excluded due to small sample size ($n \approx 60$). (w) = within-dyad effect; (b) = between-dyad effect. Stars indicate posterior credible-interval exclusion of zero: * 90\%, ** 95\%, *** 99\%. They are not p-values. Effective estimation sample: short-run N = 7,025, dyads = 403, years = 20; long-run N = 6,468, dyads = 403, years = 19.
\end{tablenotes}
\end{threeparttable}
\end{table}

Taken together, these estimates are consistent with a frequency-dependent WB convergence. Short-run synchronization remains weaker for EU--WB pairs but is more favorable when bilateral trade deepens, whereas long-run convergence is more strongly associated with reductions in structural dissimilarity within EU--WB dyads than with short-run trade fluctuations. Appendix~\ref{app:wb_prior_sensitivity} shows that the stronger WB zero-inflation interaction results are not stable across prior structures.

To examine within-WB heterogeneity, Table~\ref{tab:wb_3way} replaces the single EU--WB indicator with country-specific terms. We keep this table in the main text, but interpret it more cautiously because the country-specific WB estimates are more prior-sensitive than the pooled WB convergence results and are based on a smaller sample, which further limits precision. In the short-run band, the negative mean-gap signal is concentrated in MKD. The Panel A EU--MKD main effect in the mean equation is $-0.44$, while the corresponding EU--SRB and EU--MNE terms are not credibly different from zero. At the same time, the Panel A zero-inflation main effects have positive posterior means for all three countries and rise in magnitude from MKD ($0.62$) to SRB ($1.24$) to MNE ($2.45$). This result indicates that the evidence for higher short-run zero-state risk relative to EU--EU dyads is strongest for SRB and especially MNE, while MKD points in the same direction but is less precisely estimated.

\begin{table}[htbp]
\centering
\caption{Country heterogeneity: MKD vs.\ SRB vs.\ MNE}
\label{tab:wb_3way}
\footnotesize
\setlength{\tabcolsep}{3.5pt}
\renewcommand{\arraystretch}{0.92}
\begin{threeparttable}
\begin{tabular}{lcccccc}
\toprule
& \multicolumn{3}{c}{Short-run (1.5--4.5y)} & \multicolumn{3}{c}{Long-run (4.5--8.5y)} \\
\cmidrule(lr){2-4} \cmidrule(lr){5-7}
Coefficient & MKD & SRB & MNE & MKD & SRB & MNE \\
\midrule
\multicolumn{7}{l}{\textit{Panel A: Main effects}} \\
EU--country main effect ($\mu$) & $-$0.44*** & $-$0.12 & 0.06 & 0.10 & 0.24 & 0.41 \\
                  & {\scriptsize [-0.69, -0.21]} & {\scriptsize [-0.44, 0.21]} & {\scriptsize [-0.32, 0.46]} & {\scriptsize [-0.21, 0.40]} & {\scriptsize [-0.18, 0.67]} & {\scriptsize [-0.07, 0.91]} \\
EU--country main effect ($zi$) & 0.62* & 1.24*** & 2.45*** & 0.38 & $-$0.43 & 1.01* \\
                  & {\scriptsize [-0.01, 1.28]} & {\scriptsize [0.53, 1.95]} & {\scriptsize [1.75, 3.16]} & {\scriptsize [-0.67, 1.49]} & {\scriptsize [-1.55, 0.66]} & {\scriptsize [-0.06, 2.10]} \\
\midrule
\multicolumn{7}{l}{\textit{Panel B: Trade interactions}} \\
EU--country main effect ($\mu$) & $-$0.64*** & $-$0.11 & $-$0.01 & 0.17 & 0.29 & 0.21 \\
                  & {\scriptsize [-0.99, -0.29]} & {\scriptsize [-0.45, 0.24]} & {\scriptsize [-0.78, 0.78]} & {\scriptsize [-0.23, 0.55]} & {\scriptsize [-0.16, 0.74]} & {\scriptsize [-0.69, 1.13]} \\
EU--country main effect ($zi$) & 0.75* & 1.40*** & 2.06*** & 0.38 & $-$0.23 & 0.77 \\
                  & {\scriptsize [-0.05, 1.57]} & {\scriptsize [0.65, 2.17]} & {\scriptsize [1.01, 3.10]} & {\scriptsize [-0.75, 1.58]} & {\scriptsize [-1.39, 0.92]} & {\scriptsize [-0.47, 2.02]} \\
(W) trade intensity $\times$ EU--country ($\mu$) & 1.10** & 1.56*** & $-$0.09 & $-$0.42 & 0.27 & 0.24 \\
                  & {\scriptsize [0.08, 2.15]} & {\scriptsize [0.57, 2.56]} & {\scriptsize [-2.00, 1.77]} & {\scriptsize [-1.54, 0.71]} & {\scriptsize [-0.63, 1.19]} & {\scriptsize [-1.14, 1.71]} \\
(W) trade intensity $\times$ EU--country ($zi$) & $-$0.33 & $-$0.88 & 0.06 & -0.00 & 0.29 & $-$0.04 \\
                  & {\scriptsize [-2.14, 1.48]} & {\scriptsize [-2.57, 0.85]} & {\scriptsize [-1.82, 1.93]} & {\scriptsize [-1.88, 1.88]} & {\scriptsize [-1.55, 2.14]} & {\scriptsize [-1.99, 1.87]} \\
(B) trade intensity $\times$ EU--country ($\mu$) & $-$0.47 & 0.13 & $-$0.15 & 0.20 & 0.11 & $-$0.41 \\
                  & {\scriptsize [-1.15, 0.22]} & {\scriptsize [-0.40, 0.68]} & {\scriptsize [-1.55, 1.28]} & {\scriptsize [-0.51, 0.91]} & {\scriptsize [-0.47, 0.68]} & {\scriptsize [-1.98, 1.21]} \\
(B) trade intensity $\times$ EU--country ($zi$) & 0.31 & 0.52 & $-$0.86 & $-$0.04 & 0.97 & $-$0.53 \\
                  & {\scriptsize [-1.13, 1.77]} & {\scriptsize [-0.79, 1.83]} & {\scriptsize [-2.60, 0.88]} & {\scriptsize [-1.77, 1.65]} & {\scriptsize [-0.60, 2.57]} & {\scriptsize [-2.27, 1.25]} \\
\midrule
\multicolumn{7}{l}{\textit{Panel C: Structural interactions}} \\
EU--country main effect ($\mu$) & $-$0.50*** & $-$0.05 & $-$0.02 & 0.14 & 0.25 & 0.39 \\
                  & {\scriptsize [-0.78, -0.22]} & {\scriptsize [-0.40, 0.30]} & {\scriptsize [-0.44, 0.38]} & {\scriptsize [-0.21, 0.48]} & {\scriptsize [-0.19, 0.69]} & {\scriptsize [-0.16, 0.92]} \\
EU--country main effect ($zi$) & 0.68* & 1.23*** & 2.35*** & 0.44 & $-$0.27 & 0.99 \\
                  & {\scriptsize [-0.02, 1.42]} & {\scriptsize [0.49, 1.97]} & {\scriptsize [1.55, 3.15]} & {\scriptsize [-0.73, 1.58]} & {\scriptsize [-1.39, 0.86]} & {\scriptsize [-0.19, 2.17]} \\
(W) specialization distance $\times$ EU--country ($\mu$) & $-$0.16 & 0.23 & 0.19 & $-$0.22 & $-$0.08 & $-$0.77*** \\
                  & {\scriptsize [-0.44, 0.12]} & {\scriptsize [-0.39, 0.83]} & {\scriptsize [-0.37, 0.78]} & {\scriptsize [-0.55, 0.11]} & {\scriptsize [-0.53, 0.37]} & {\scriptsize [-1.33, -0.21]} \\
(W) specialization distance $\times$ EU--country ($zi$) & $-$0.23 & $-$0.88 & $-$0.37 & $-$0.07 & 0.28 & $-$0.87 \\
                  & {\scriptsize [-0.95, 0.48]} & {\scriptsize [-1.99, 0.27]} & {\scriptsize [-1.37, 0.63]} & {\scriptsize [-1.07, 1.00]} & {\scriptsize [-1.00, 1.61]} & {\scriptsize [-2.16, 0.38]} \\
(B) specialization distance $\times$ EU--country ($\mu$) & 0.15 & $-$0.05 & 0.38 & $-$0.12 & 0.01 & 0.11 \\
                  & {\scriptsize [-0.18, 0.49]} & {\scriptsize [-0.32, 0.24]} & {\scriptsize [-0.19, 0.96]} & {\scriptsize [-0.63, 0.38]} & {\scriptsize [-0.31, 0.33]} & {\scriptsize [-0.49, 0.70]} \\
(B) specialization distance $\times$ EU--country ($zi$) & $-$0.18 & 0.02 & 0.51 & $-$0.37 & $-$0.60 & $-$0.03 \\
                  & {\scriptsize [-1.08, 0.67]} & {\scriptsize [-0.69, 0.70]} & {\scriptsize [-0.74, 1.74]} & {\scriptsize [-1.76, 1.03]} & {\scriptsize [-1.65, 0.50]} & {\scriptsize [-1.56, 1.55]} \\
\bottomrule
\end{tabular}
\begin{tablenotes}
\scriptsize
\item Notes: Posterior means with 95\% credible intervals in brackets. The table reports country main effects and country-specific within/between interactions for trade intensity and specialization distance. All specifications are estimated with controls: lagged dependent variable, main effects for trade intensity and specialization distance (within and between), fiscal-gap, urban-gap, remittance-gap, inflation-gap, and financial-openness terms in the mean equation, plus year fixed effects and dyad random intercepts; the zero-inflation equation includes lagged dependent variable, trade, specialization distance, financial openness, year fixed effects, and dyad random intercepts. EU-MKD, EU-SRB, EU-MNE = pairs involving one EU member and one WB country. Stars indicate posterior credible-interval exclusion of zero: * 90\%, ** 95\%, *** 99\%. They are not p-values. $\mu$ = mean equation; $zi$ = zero-inflation equation. (W) = deviation from a dyad's own historical mean; (B) = a dyad's long-run average relative to other dyads. Effective estimation sample: short-run N = 7,025, dyads = 403, EU-country dyads per WB country = 26 each; long-run N = 6,468, dyads = 403, EU-country dyads per WB country = 26 each.
\end{tablenotes}
\end{threeparttable}
\end{table}

Trade-related heterogeneity is strongest for MKD and SRB. In the short-run mean equation, the within-dyad trade interactions are positive for EU--MKD ($1.10$) and EU--SRB ($1.56$), but not for EU--MNE. Under the intermediate priors, however, the corresponding zero-inflation trade interactions are not credibly different from zero. In the long-run band, trade interactions are generally imprecise across all three countries. For Serbia, the long-run between-dyad trade interaction in the zero-inflation equation is not credibly different from zero under this specification. This suggests that the country-specific trade evidence is concentrated in the short-run conditional mean rather than in the probability of falling into the zero-synchronization regime.

Structural heterogeneity appears primarily in MNE at long horizons. The long-run within-dyad specialization-distance interaction for EU--MNE is negative in the mean equation ($-0.77$), while analogous terms for MKD and SRB remain close to zero. Under the prior used here, the corresponding zero-inflation structural interactions are not credibly different from zero. These country-level results should be interpreted with caution, because subgroup estimates are noisier than pooled estimates and several intervals remain wide.

Overall, the WB evidence does not point to a single convergence mechanism across countries. The short-run trade association is most pronounced for MKD and SRB, while the long-run role of reduced structural dissimilarity is most visible for MNE. At the same time, the country-specific zero-inflation results are more prior-sensitive than the mean-equation results, so we treat the latter as the core WB heterogeneity findings and the former as exploratory. Appendix~\ref{app:wb_prior_sensitivity} reports how the regularizing and default-prior alternatives change the strength of the WB three-way evidence. This mixed pattern is compatible with distinct integration paths across WB economies and reinforces the broader conclusion that convergence channels are frequency dependent and heterogeneous.

\begin{table}[htbp]
\centering
\caption{Comparison of results with the related literature}
\label{tab:literature_findings_comparison}
\scriptsize
\renewcommand{\arraystretch}{1.15}
\begin{threeparttable}
\resizebox{\textwidth}{!}{%
\begin{tabular}{p{3.1cm}p{5.2cm}p{5.4cm}p{2.4cm}p{3.4cm}}
\toprule
\textbf{Finding} & \textbf{Description} & \textbf{Closest literature benchmark} & \textbf{Comparison} & \textbf{References} \\
\midrule
\multicolumn{5}{l}{\textit{Full-sample findings}} \\
\addlinespace[0.4em]
Frequency dependence of synchronization & The sign and strength of determinants differ across the 1.5--4.5 year and 4.5--8.5 year bands. & Time--frequency studies show that European synchronization is heterogeneous, time-varying, and frequency-dependent, but they generally do not estimate frequency-specific determinants. & Directly consistent & \cite{soares2011business,crowley2005decomposing,bekiros2015business,njegic2017business} \\
\addlinespace[0.3em]
Institutional integration in the short run & Joint EU and EMU membership is positively associated with short-run synchronization and with a lower probability of zero synchronization. & The standard endogeneity literature and several CESEE studies associate deeper EU/EMU integration with stronger cyclical alignment. & Directly consistent & \cite{frankel1998endogenity,darvas2008business,botric2016exploring} \\
\addlinespace[0.3em]
Institutional integration in the long run & Within-dyad EU/EMU effects reverse sign at longer frequencies, while between-dyad effects become negligible. & The average endogeneity result predicts non-negative integration effects, but specialization-based arguments and shock-type distinctions imply that deeper integration can generate divergence at longer horizons or in supply-side dimensions. & Indirect/theoretical support only & \cite{Krugman1993,fidrmuc2004endogeneity,babetskii2005trade} \\
\addlinespace[0.3em]
Trade channel & Trade matters at both horizons, but through different margins: short-run synchronization is mainly associated with between-dyad differences in average trade intensity, while long-run synchronization is associated more with within-dyad trade deepening. & The literature usually finds a positive average trade--synchronization link, especially when integration raises intra-industry trade, but also stresses that the mechanism depends on trade composition and the type of shocks transmitted. & Partial direct support & \cite{frankel1998endogenity,fidrmuc2004endogeneity,imbs2004trade,babetskii2005trade} \\
\addlinespace[0.3em]
Structural similarity / specialization & Specialization distance is negatively associated with synchronization in both bands, with the long-run association concentrated especially in between-dyad structural differences and in the within-dyad zero-inflation margin. & Similar production structures are generally associated with stronger co-movement, while specialization can increase exposure to asymmetric shocks and weaken synchronization. & Partial direct support & \cite{imbs2004trade,Krugman1993,fidrmuc2004endogeneity} \\
\addlinespace[0.3em]
Financial openness & Financial openness shows mixed short-run evidence across margins, but is negatively associated with long-run synchronization. & WB-specific evidence often finds financial and remittance channels to be de-synchronizing, whereas broader European evidence is mixed and sometimes points in the opposite direction. & Mixed direct evidence & \cite{hildebrandt2015business,imbs2004trade,botric2019business} \\
\midrule
\multicolumn{5}{l}{\textit{Western Balkans pooled convergence findings}} \\
\addlinespace[0.4em]
Short-run EU--WB gap and trade channel & EU--WB dyads are less synchronized than EU--EU dyads in the short run, and trade deepening within EU--WB dyads is more positively associated with synchronization than in EU--EU dyads in the conditional mean; the stronger zero-state-risk effect is most visible under the default-prior WB specification and weakens under the more regularizing alternatives. & Earlier WB and SEE studies generally find that WB/SEE economies are less synchronized with the euro area than more integrated peers, and that trade is the main positive convergence channel. & Directly consistent in the mean equation; evidence on the probability of falling into the zero-synchronization regime is prior-sensitive & \cite{hildebrandt2015business,Gouveia2014,botric2016bilateral,kovavcic2017assessing} \\
\addlinespace[0.3em]
Long-run EU--WB convergence via structural alignment & At longer frequencies the average EU--WB gap largely disappears, while structural similarity matters more than trade in the conditional-mean equation; the corresponding zero-inflation structural terms are not robust across prior structures. & Earlier WB studies rely mostly on time-domain measures and do not cleanly separate long-run convergence from short-run co-movement, so the structural long-run mechanism is largely unresolved. & Indirect/theoretical support only & \cite{hildebrandt2015business,imbs2004trade,Krugman1993} \\
\midrule
\multicolumn{5}{l}{\textit{WB country heterogeneity}} \\
\addlinespace[0.4em]
North Macedonia (MKD) & MKD shows the clearest short-run mean-gap signal relative to EU--EU dyads, but also a strong positive short-run trade interaction in the mean equation and no clear long-run average gap. & Macedonia is described as strongly and increasingly synchronized with the Euro-zone, with Euro-zone shocks transmitted quickly and persistently; newer IMF evidence also finds the strongest euro-area real-shock exposure in the region because of deep trade and GVC linkages. & Mixed direct evidence & \cite{filipovski2018business,Gouveia2014,ma2025impact} \\
\addlinespace[0.3em]
Serbia (SRB) & SRB shows elevated short-run zero-state risk and positive short-run trade interactions in the mean equation, while long-run trade effects are imprecise. & Serbia is usually described as less synchronized than several regional peers, with only partial convergence over time and no robust trade--synchronization relationship in small-sample correlation studies. & Mixed direct evidence & \cite{obradovic2013synchronization,Gouveia2014,kovavcic2017assessing} \\
\addlinespace[0.3em]
Montenegro (MNE) & MNE shows the highest short-run zero-state risk, no clear short-run trade effect, and the strongest long-run structural interaction in the mean equation. & Montenegro shows some positive real-activity synchronization with the euro area, and euroization appears to import broadly countercyclical output stabilization from ECB policy, but monetary transmission remains highly domestic and idiosyncratic. Broader WB evidence still suggests relatively stronger later synchronization than for Serbia. & Indirect/theoretical support only & \cite{hildebrandt2015business,sokic2013euroization,ma2025impact} \\
\bottomrule
\end{tabular}%
}
\begin{minipage}{0.98\textwidth}
\footnotesize
\textit{Notes}: The table focuses on the main findings emphasized in the results section. The comparison labels distinguish close empirical matches from partial, mixed, or mainly theoretical support. For the WB rows, the descriptions refer to the intermediate-prior specification used in the main text unless otherwise noted.
\end{minipage}
\end{threeparttable}
\end{table}

\section{Conclusion}\label{sec:conclusion}

In this paper we analyzed business-cycle synchronization between EU member states and Western Balkan candidates by combining wavelet-based time--frequency measures with Bayesian zero-inflated beta panel models that distinguish short-run (1.5--4.5 years) from long-run (4.5--8.5 years) co-movement. The empirical design focused on annual dyad-year synchronization outcomes and decomposed the main covariates into within- and between-dyad components, allowing us to ask whether the correlates of synchronization differ across time horizons and whether the Western Balkans converge toward EU through the same channels as the broader European sample.

The results point to a clear frequency dependence. At shorter horizons, synchronization is highly persistent and is positively associated with shared EU and EMU membership, while trade matters mainly through cross-dyad differences in average bilateral intensity. On the other hand, at longer horizon the within-dyad EU and EMU coefficients turn negative or lose precision, trade deepening within a dyad becomes a stronger correlate of synchronization, and lower between-dyad specialization distance becomes more important as a correlate of synchronization. The Western Balkans results fit this broader pattern. EU--WB dyads remain less synchronized than EU--EU pairs in the short run, but trade deepening within those dyads is more positively associated with short-run co-movement than in EU--EU dyads; at longer horizons, the average EU--WB gap narrows, and the remaining long-run convergence result is more consistent with sectoral differences within dyads declining over time than with a separate positive effect of lower structural dissimilarity in levels. However, the WB zero-inflation interaction results are noticeably more prior-sensitive than the corresponding mean-equation results, especially in the country-specific analysis; see Appendix~\ref{app:wb_prior_sensitivity}.

These findings suggest that institutional integration and market integration need not operate through a single mechanism. Nevertheless, one interpretation of the evidence connects to a longstanding debate in the optimal currency area literature. Frankel and Rose~\cite{frankel1998endogenity} argued that monetary integration endogenously promotes synchronization through increased trade and policy convergence, while Krugman~\cite{Krugman1993} emphasized that integration may encourage specialization along comparative advantage, ultimately making economic structures more heterogeneous. Our estimates are consistent with both mechanisms operating simultaneously but at different time scales. At shorter horizons, institutional integration is associated with higher synchronization, which is what one would expect if common monetary policy and reduced trade barriers help transmit short-run demand shocks across borders. At longer horizons, by contrast, the negative within-dyad EU and EMU associations, together with the increased importance of specialization distance, are consistent with the Krugman view that deeper integration may foster structural divergence and contribute to weaker co-movement. This comparison with the literature is summarized in Table~\ref{tab:literature_findings_comparison}, which shows that the short-run institutional and trade results are directly more consistent with existing evidence, whereas the long-run negative or null institutional effects receive mainly indirect support from the specialization literature rather than from close empirical counterparts.

However, these results should be interpreted cautiously. Reverse causality remains possible, and omitted factors such as geographic proximity, shared history, or common legal traditions may still account for part of the estimated associations. The specification also imposes strong assumptions: the REWB decomposition treats within- and between-dyad effects as linear and additive, year effects absorb common shocks but may also absorb economically meaningful EU-wide policy responses during crises, and the inclusion of lagged synchronization with dyad effects may introduce modest finite-sample bias. In addition, the zero-inflation component assumes that observed zeros reflect genuine absence of synchronization rather than measurement error or low-power wavelet detection in noisy dyad-years. In addition, the results should be read as conditional associations consistent with theoretical mechanisms rather than causal estimates. Future work should therefore test nonlinear and interactive specifications more systematically, examine bias-correction or instrumental-variable approaches for the dynamic panel structure, validate the zero-state classification with alternative measurement strategies, and pursue stronger identification through accession-timing designs, policy discontinuities, or richer sector-level data.

\appendix

\section*{Appendix}
\section{Wavelet Analysis: Mathematical Details}
\label{app:wavelet_math}

This appendix provides mathematical details for the wavelet coherence methodology.

Wavelet analysis rests on the continuous wavelet transform (CWT), which projects a time series onto a family of localized basis functions~\cite{torrence1998practical,aguiar2014continuous}. Let $x(t)$ be a square-integrable time series and let $\psi(t)$ be an admissible mother wavelet satisfying certain regularity conditions (zero mean, square integrability, and the admissibility constant being finite). The CWT of $x$ at scale $s>0$ and location $\tau\in\mathbb{R}$ is defined as the inner product
\begin{equation}
W_x(s,\tau)
=\big\langle x,\psi_{s,\tau}\big\rangle
=\int_{-\infty}^{\infty} x(t)\,\frac{1}{\sqrt{|s|}}\,
\psi^{\!*}\!\left(\frac{t-\tau}{s}\right)\,dt,
\label{eq:cwt}
\end{equation}
where ${}^{*}$ denotes complex conjugation and $\psi_{s,\tau}(t)=|s|^{-1/2}\,\psi\!\left(\frac{t-\tau}{s}\right)$ is the scaled and translated wavelet. The $1/\sqrt{|s|}$ normalization ensures energy preservation across scales. For complex (analytic) wavelets such as the Morlet, $W_x$ is complex-valued and we can write its amplitude $|W_x|$ and phase $\phi_x(s,\tau)=\arctan\!\big(\Im W_x/\Re W_x\big)$. The analytic property is crucial for extracting phase relationships between time series.

We use the Morlet wavelet throughout this analysis, defined as a complex sine wave modulated by a Gaussian envelope~\cite{torrence1998practical,aguiar2014continuous}. With central frequency $\mu_f=\eta/2\pi$ (typically $\eta=6$ to satisfy the admissibility condition), the Fourier frequency associated with a scale $s$ follows the linear relationship
\begin{equation}
f \;=\; \frac{\mu_f}{s}, \qquad \text{and hence period } P=\frac{1}{f}=\frac{s}{\mu_f}.
\label{eq:scale_freq}
\end{equation}
This scale-to-frequency mapping allows us to interpret wavelet coefficients in terms of familiar periodicities. For example, to isolate business cycles of 2--8 years, we restrict analysis to scales corresponding to those periods via equation \eqref{eq:scale_freq}~\cite{soares2011business}.

The local wavelet power spectrum of $x$ is defined as the squared modulus of the wavelet transform,
\begin{equation}
S_x(s,\tau)\;=\;|W_x(s,\tau)|^2,
\label{eq:wps}
\end{equation}
which measures the local contribution to variance at time-scale location $(s,\tau)$. This is the wavelet analogue of the power spectral density in Fourier analysis, but localized in time. Statistical significance of $S_x$ is typically assessed by comparing it against a background spectrum. Under the null hypothesis of white noise or AR(1) red noise, the expected distribution of $|W_x|^2$ can be derived analytically~\cite{torrence1998practical}. We use an alternative approach: phase-randomized Fourier surrogates, which preserve the Fourier power spectrum, and hence the linear autocorrelation structure, of the original series while randomizing phases~\cite{theiler1992testing,schreiber2000surrogate}. By generating an ensemble of surrogate time series and computing their wavelet transforms, we construct an empirical null distribution against which observed power is tested. Relative to a specific white- or red-noise benchmark, this yields a spectrum-matched empirical null. However, pure phase randomization does not generally preserve the marginal amplitude distribution exactly, so the interpretation is a test against a linear spectrum-preserving null rather than against every possible non-Gaussian linear alternative~\cite{theiler1992testing,schreiber2000surrogate}.

To measure co-movement between two time series $x(t)$ and $y(t)$, we begin with their respective wavelet transforms $W_x$ and $W_y$. The cross-wavelet transform is the product of one transform with the complex conjugate of the other,
\begin{equation}
W_{xy}(s,\tau) \;=\; W_x(s,\tau)\,W_y^{\!*}(s,\tau),
\label{eq:cross}
\end{equation}
and its modulus $|W_{xy}(s,\tau)|$ gives the cross-wavelet power. However, raw cross-wavelet power is not normalized and depends on the individual series' magnitudes. To obtain a localized correlation measure analogous to the Pearson coefficient, we compute wavelet coherency. This requires smoothing both the cross-wavelet and the individual power spectra to ensure statistical stability. Denote this smoothing operator by $S(\cdot)$, which typically involves a Gaussian window in time and a box (rectangular) window in scale. The wavelet coherency is then
\begin{equation}
R_{xy}(s,\tau)
\;=\;
\frac{\big|\,S\!\big(W_{xy}(s,\tau)\big)\big|}
{\big(S(|W_x(s,\tau)|^2)\big)^{1/2}\,
 \big(S(|W_y(s,\tau)|^2)\big)^{1/2}}.
\label{eq:coherency_app}
\end{equation}
This quantity ranges from 0 (no coherence) to 1 (perfect coherence) and measures the localized correlation between $x$ and $y$ in the time-frequency domain. The smoothing is essential: without it, coherency would be identically one at all time-scale points due to the degrees of freedom being exhausted by the normalization~\cite{grinsted2004application,aguiar2014continuous}.

Beyond measuring the strength of co-movement, we can extract directional information from the phase of the cross-wavelet. The phase difference between $x$ and $y$ at $(s,\tau)$ is
\begin{equation}
\phi_{x,y}(s,\tau)
\;=\;
\arctan\!\left(
\frac{\Im\{W_{xy}(s,\tau)\}}
     {\Re\{W_{xy}(s,\tau)\}}
\right)
\in[-\pi,\pi].
\label{eq:phase_diff_app}
\end{equation}
The interpretation is as follows. When $\phi_{x,y}=0$, the two series are in phase (peaks and troughs coincide). On the other hand, when $\phi_{x,y}=\pi$ or $-\pi$, they are in anti-phase (one peaks when the other troughs). For intermediate values, one series leads or lags the other: $\phi_{x,y}\in(0,\pi/2)$ means $y$ leads $x$ at that time-scale location, while $\phi_{x,y}\in(-\pi/2,0)$ means $x$ leads $y$~\cite{grinsted2004application,aguiar2014continuous}. This phase information is particularly valuable when studying transmission of shocks between economies, as it reveals which country's cycle tends to precede the other's.

For reporting purposes, it is often useful to convert phase differences into time lags~\cite{aguiar2014continuous}. Within a given frequency band $[f_1,f_2]$ (equivalently, a scale or period band via equation \eqref{eq:scale_freq}), we compute the instantaneous time lag at location $\tau$ as
\begin{equation}
\Delta T(\tau)
\;=\;
\frac{\phi_{x,y}(\tau)}{2\pi\,F(\tau)}\,,
\qquad
F(\tau)
=
\frac{\displaystyle \int_{f_1}^{f_2} f\,|W_{xy}(f,\tau)|\,df}
     {\displaystyle \int_{f_1}^{f_2}    |W_{xy}(f,\tau)|\,df},
\label{eq:time_lag_app}
\end{equation}
where $F(\tau)$ is the first normalized spectral moment of the cross-wavelet power over the band, representing the dominant frequency at time $\tau$. The time lag $\Delta T(\tau)$ is then measured in the same units as the original time series (months in our case). A positive lag indicates that $y$ leads $x$; a negative lag indicates that $x$ leads $y$. These lead-lag relationships can vary over time and across frequency bands, reflecting the time-varying and scale-dependent nature of economic co-movement. While we do not model $\Delta T$ directly in our panel regression (we focus on unsigned coherence), we report it descriptively in Figure~\ref{fig:wavelet_grid} to characterize the directional dynamics between Western Balkan countries and the EU aggregate.

\section{Data Construction}
\label{app:data}

The construction of our annual synchronization measure requires several preprocessing steps that balance statistical reliability against sample size. Two aspects require detailed explanation: the cone of influence filtering and the band-averaging weights.

Wavelet coefficients near the boundaries of a time series are contaminated by edge effects. At the beginning and end of the sample, the wavelet window extends beyond the available data, and the transform must be computed using implicit assumptions (typically zero-padding or periodic extension). These boundary regions constitute the cone of influence (COI), within which coefficients are unreliable. The spatial extent of the COI depends on the scale, thus longer-period wavelets have wider windows and thus larger edge regions. For a given dyad-year, some months may fall entirely within the COI for certain scales, while other months are sufficiently far from the edges to yield reliable estimates.

We apply a two-stage filtering strategy. First, we exclude any dyad-year with fewer than nine total months of data in that calendar year. This ensures a minimum temporal coverage for annual aggregation. Second, within the remaining dyad-years, we exclude any month that falls within the COI for the frequency band of interest. For the short-run band (1.5--4.5 years), the COI boundary is determined by the longest period (4.5 years); for the long-run band (4.5--8.5 years), it is determined by the 8.5-year period. We then drop dyad-years where fewer than six eligible months remain outside the COI. This threshold balances two concerns. Overly tolerant filtering (e.g., requiring only three eligible months) would introduce spurious zeros from low coverage: pairs with only a few usable months might register zero synchronization simply because the available data are sparse, not because the economies are genuinely desynchronized. On the other hand, overly stringent filtering (e.g., requiring all twelve months) would discard too many observations, particularly for dyad-years at the beginning and end of the sample period. The six-month threshold ensures that each annual measure is based on at least half a year of reliable monthly estimates, providing a reasonable balance between coverage and data quality.

Importantly, this filtering removes measurement zeros (zeros arising from insufficient data) but retains structural zeros (dyad-years where the economies genuinely did not exhibit significant coherence at the relevant frequencies). The zero-inflated component of our regression model is designed to accommodate these structural zeros, interpreting them as informative outcomes rather than missing data.

The second technical detail concerns the band-averaging weights. When aggregating coherence across scales within a frequency band, a naive unweighted average would give disproportionate influence to longer-period components. Consider the short-run band spanning 1.5 to 4.5 years. A cycle with period 4.5 years completes one full oscillation every 4.5 years, while a cycle with period 1.5 years completes three oscillations in the same timespan. If we average coherence values across all scales in the band with equal weights, we are effectively averaging over different amounts of information: the 4.5-year component contributes fewer independent observations (fewer complete cycles) than the 1.5-year component. To correct for this, we weight each scale $s$ by the inverse of its period $P_s$, so that $w_s \propto 1/P_s$ with $\sum_s w_s = 1$. This ensures that the band-averaged coherence reflects a frequency-balanced summary rather than being dominated by long-period scales. Without this correction, coherence estimates would be biased toward lower-frequency components simply due to their temporal span. The $1/P_s$ weighting down-weights longer periods proportionally, so that all frequency components within the band contribute comparably to the final average. This is analogous to spectral averaging with logarithmic frequency bins in traditional Fourier analysis, where frequency ranges are weighted proportionally.

Together, these preprocessing choices ensure that our annual synchronization measure is both statistically reliable (free from edge effects and dominated by well-identified coherence) and methodologically sound (appropriately aggregated across the frequency band). The resulting dependent variable captures genuine time-varying co-movement at specific cyclical frequencies, providing a suitable foundation for the panel regression analysis presented in the main text.

We next explain the preprocessing applied to the covariates before model estimation. Variables reported in percentage units are converted to ratios when their empirical scale indicates percentage encoding, and sectoral shares (agriculture, industry, services) are renormalized so they sum to one. We then construct transformed country-level inputs used in pairwise gaps, including real capital per capita, log human capital, and external-balance ratios used in financial openness, to make explicit that dyadic regressors are built from these derived country-level measures rather than directly from raw series.
Missing covariate values are handled with within-country linear interpolation for internal gaps up to five years, without extrapolation outside observed endpoints; The country-year panel is then converted to undirected dyad-years and the bilateral gap regressors are formed as absolute cross-country differences. Finally, before model fitting, continuous regressors are standardized and decomposed into within- and between-dyad components (REWB), and the first lagged dependent-variable observation per dyad is dropped.

Note that, liquid liabilities gap, financial system deposits gap, and bank deposits gap were excluded by the pre-specified collinearity filter because they were highly correlated (absolute pairwise correlation above $0.85$), which would otherwise lead to unstable and weakly identified coefficient estimates.

\section{Interpreting the REWB Decomposition: An Illustrative Example}\label{app:illustrative-example}

To illustrate the REWB interpretation, consider a hypothetical covariate $x_{gt}$ (e.g., trade intensity). Suppose the total mean is $\bar{x}=6$. Dyad A has a sample-period mean of $\bar{x}_{A}=8$, while Dyad B has $\bar{x}_{B}=4$. In one year, both dyads are observed one unit above their own sample-period means ($x_{A,t}=9$, $x_{B,t}=5$), so both have the same within component: $x^w_{A,t}=x^w_{B,t}=1$. Their between components differ, however: $x^b_{A,t}=8-6=2$ and $x^b_{B,t}=4-6=-2$.

If the mean equation has hypothetical coefficients $\beta_{\mu}^{w}=0.10$ and $\beta_{\mu}^{b}=0.25$, the contribution of $x$ to the linear predictor is
\begin{equation}
\Delta \eta^{(\mu)}_{A,t}=0.10\cdot 1+0.25\cdot 2=0.60, \qquad
\Delta \eta^{(\mu)}_{B,t}=0.10\cdot 1+0.25\cdot (-2)=-0.40.
\end{equation}
The positive value for Dyad A (+0.60) means that, in this year, the trade variable pushes A's latent mean-synchronization index\footnote{Index refers to the logit of the conditional mean.} upward; the negative value for Dyad B ($-$0.40) means the same variable pushes B's index downward. Importantly, both dyads have the same within deviation ($x^w=1$), so the difference comes entirely from their between positions. A has a high dyad mean relative to the sample ($x^b=2$), while B has a low dyad mean ($x^b=-2$)\footnote{These are changes in the linear predictor (logit scale), not direct percentage-point changes in synchronization.}. This simple example shows the distinction used throughout the results: the within effect ($\beta_{\mu}^{w}$) captures how synchronization changes when a given country pair is above or below its own sample-period mean in a particular year, while the between effect ($\beta_{\mu}^{b}$) captures how country pairs with higher or lower sample-period means differ from other pairs on average.

For institutional variables, define
\begin{equation}
B^{EU}_{gt}=\mathbf{1}\{a\in EU_t \ \text{and}\ b\in EU_t\}, \qquad
B^{EMU}_{gt}=\mathbf{1}\{a\in EMU_t \ \text{and}\ b\in EMU_t\},
\end{equation}
which correspond to \texttt{both\_eu} and \texttt{both\_emu}. Their REWB components are
\begin{equation}
\left(B^{EU}_{gt}\right)^w = B^{EU}_{gt}-\overline{B}^{EU}_g,\quad
\left(B^{EU}_{gt}\right)^b = \overline{B}^{EU}_g-\overline{B}^{EU},
\end{equation}
and analogously for $B^{EMU}_{gt}$.

Consider the sample period 2001--2021. For Croatia--Slovenia, $B^{EU}_{gt}=0$ in 2001--2012 and $B^{EU}_{gt}=1$ in 2013--2021, so $\overline{B}^{EU}_g=9/21=0.429$. In 2011 (before both are in the EU), $\left(B^{EU}_{gt}\right)^w=0-0.429=-0.429$; in 2015 (after both are in the EU), $\left(B^{EU}_{gt}\right)^w=1-0.429=0.571$. This is what the within coefficient for EU uses: a within-dyad before/after contrast around that dyad's own sample-period mean. For the same dyad, $B^{EMU}_{gt}=0$ throughout 2001--2021 (Croatia joins EMU only in 2023), so $\left(B^{EMU}_{gt}\right)^w=0$ in every year; this dyad contributes no within variation for EMU in our estimation window.

Now compare across dyads. Germany--France has $B^{EU}_{gt}=1$ and $B^{EMU}_{gt}=1$ in every year, so $\overline{B}^{EU}_g=\overline{B}^{EMU}_g=1$. If, illustratively, the panel-wide averages are $\overline{B}^{EU}=0.40$ and $\overline{B}^{EMU}=0.25$, then Germany--France has $\left(B^{EU}_{gt}\right)^b=1-0.40=0.60$ and $\left(B^{EMU}_{gt}\right)^b=1-0.25=0.75$, while Croatia--Slovenia has $\left(B^{EU}_{gt}\right)^b=0.429-0.40=0.029$ and $\left(B^{EMU}_{gt}\right)^b=0-0.25=-0.25$. This is what the between coefficients use: differences in average EU/EMU exposure across dyads.

To make the coefficient interpretation explicit, use a purely illustrative example with $\beta_{\mu,EU}^{w}=0.30$ and $\beta_{\mu,EU}^{b}=0.40$. For Croatia--Slovenia, $B^{EU}_{gt}=0$ before 2013 and $B^{EU}_{gt}=1$ from 2013 onward, so $\overline{B}^{EU}_g=9/21=0.429$. In 2011, $\left(B^{EU}_{gt}\right)^w=-0.429$; in 2015, $\left(B^{EU}_{gt}\right)^w=0.571$. The within-EU contribution to the mean index is therefore $0.30\times(-0.429)=-0.129$ in 2011 and $0.30\times0.571=0.171$ in 2015. The accession-related change is $+0.300$ (because the within term increases by exactly 1 when $B^{EU}_{gt}$ switches from 0 to 1), meaning EU accession pushes this dyad's latent synchronization index upward in this example.

For the between effect, keep $\overline{B}^{EU}=0.40$ as the panel mean. Croatia--Slovenia has $\left(B^{EU}_{gt}\right)^b=0.429-0.40=0.029$, while Germany--France (always jointly in the EU) has $\left(B^{EU}_{gt}\right)^b=1-0.40=0.60$. Their between-EU contributions are $0.40\times0.029=0.012$ and $0.40\times0.60=0.240$, respectively. Thus Germany--France receives a larger positive between-EU contribution by $0.240-0.012=0.228$, which is exactly what a positive between coefficient means: dyads with higher average EU exposure are predicted to have higher mean synchronization, all else equal.

The same mechanics apply in the zero-inflation equation. If, illustratively, $\beta_{\pi,EU}^{w}=-0.50$, the same 0-to-1 accession switch lowers the zero-state index by 0.50 and multiplies the odds of complete desynchronization by $\exp(-0.50)=0.61$ (a 39\% reduction in odds).

\section{Prior Sensitivity of the Western Balkans Results}
\label{app:wb_prior_sensitivity}

Table~\ref{tab:prior_structures} in the main text summarizes the three prior regimes used in the paper. The WB results discussed in the main text are based on the intermediate prior specification, which applies $\mathcal{N}(0,1)$ priors to the population-level coefficients. This appendix reports the same WB specifications under the two alternative prior structures: a more regularized specification based on the baseline prior structure used in the main European tables, and a less regularized specification based on the default \texttt{brms} setup, under which the main population-level coefficients are estimated without coefficient shrinkage. This comparison is useful because the WB models contain many interaction terms and are estimated on much thinner effective samples than the full European panel, especially in the country-specific three-way analysis.

Table~\ref{tab:wb_convergence_prior_comparison} reports the pooled EU--WB convergence specification under these two alternative prior structures. The main qualitative result is fairly stable across priors in the mean equation. Under both alternatives, EU--WB dyads remain less synchronized than EU--EU dyads in the short run, short-run trade deepening within EU--WB dyads is more positively associated with synchronization than in EU--EU dyads, and long-run synchronization is more clearly related to reductions in structural dissimilarity within the same dyad over time than to trade interactions. What changes more substantially is the zero-inflation equation: under the less regularized specification, the estimated effects on the probability of falling into the zero-synchronization regime are stronger, whereas under the more regularized specification these coefficients are attenuated. This is why the main text emphasizes the pooled WB mean-equation results more heavily than the corresponding zero-inflation interactions.

\begin{table}[htbp]
\centering
\footnotesize
\setlength{\tabcolsep}{4pt}
\renewcommand{\arraystretch}{0.90}
\caption{Pooled EU--WB convergence results under alternative prior structures}
\label{tab:wb_convergence_prior_comparison}
\begin{threeparttable}
\begin{tabular}{p{5.8cm}cccc}
\toprule
& \multicolumn{2}{c}{Short-run (1.5--4.5y)} & \multicolumn{2}{c}{Long-run (4.5--8.5y)} \\
\cmidrule(lr){2-3} \cmidrule(lr){4-5}
Coefficient & Estimate & 95\% CI & Estimate & 95\% CI \\
\midrule
\multicolumn{5}{l}{\textbf{Prior structure without regularization}} \\
\multicolumn{5}{l}{\textit{Panel A: Mean equation ($\mu$)}} \\[2pt]
EU-WB main effect & $-$0.63*** & {\scriptsize [$-$0.90, $-$0.36]} & 0.09 & {\scriptsize [$-$0.24, 0.42]} \\
Trade intensity (w) $\times$ EU-WB & 1.63*** & {\scriptsize [0.82, 2.52]} & 0.00 & {\scriptsize [$-$0.75, 0.76]} \\
Trade intensity (b) $\times$ EU-WB & $-$0.08 & {\scriptsize [$-$0.59, 0.46]} & 0.01 & {\scriptsize [$-$0.63, 0.67]} \\
Spec.\ distance (w) $\times$ EU-WB & 0.02 & {\scriptsize [$-$0.21, 0.26]} & $-$0.26** & {\scriptsize [$-$0.51, $-$0.01]} \\
Spec.\ distance (b) $\times$ EU-WB & 0.07 & {\scriptsize [$-$0.17, 0.31]} & 0.06 & {\scriptsize [$-$0.27, 0.39]} \\
\multicolumn{5}{l}{\textit{Panel B: Zero-inflation equation ($zi$)}} \\[2pt]
EU-WB main effect & 1.92*** & {\scriptsize [1.20, 2.67]} & 0.75 & {\scriptsize [$-$0.49, 2.01]} \\
Trade intensity (w) $\times$ EU-WB & $-$2.86** & {\scriptsize [$-$5.67, $-$0.15]} & 1.10 & {\scriptsize [$-$3.10, 5.45]} \\
Trade intensity (b) $\times$ EU-WB & 0.55 & {\scriptsize [$-$0.96, 2.07]} & 0.01 & {\scriptsize [$-$2.66, 2.65]} \\
Spec.\ distance (w) $\times$ EU-WB & $-$0.57* & {\scriptsize [$-$1.17, 0.05]} & $-$0.33 & {\scriptsize [$-$1.22, 0.56]} \\
Spec.\ distance (b) $\times$ EU-WB & $-$0.13 & {\scriptsize [$-$0.81, 0.56]} & $-$1.03* & {\scriptsize [$-$2.24, 0.16]} \\
\midrule
\multicolumn{5}{l}{\textbf{Prior structure with regularization}} \\
\multicolumn{5}{l}{\textit{Panel A: Mean equation ($\mu$)}} \\[2pt]
EU-WB main effect & $-$0.50*** & {\scriptsize [$-$0.73, $-$0.28]} & 0.11 & {\scriptsize [$-$0.14, 0.37]} \\
Trade intensity (w) $\times$ EU-WB & 0.93*** & {\scriptsize [0.28, 1.57]} & 0.04 & {\scriptsize [$-$0.56, 0.64]} \\
Trade intensity (b) $\times$ EU-WB & 0.02 & {\scriptsize [$-$0.37, 0.41]} & 0.01 & {\scriptsize [$-$0.44, 0.42]} \\
Spec.\ distance (w) $\times$ EU-WB & 0.00 & {\scriptsize [$-$0.22, 0.23]} & $-$0.25** & {\scriptsize [$-$0.50, $-$0.01]} \\
Spec.\ distance (b) $\times$ EU-WB & 0.03 & {\scriptsize [$-$0.18, 0.24]} & $-$0.06 & {\scriptsize [$-$0.32, 0.20]} \\
\multicolumn{5}{l}{\textit{Panel B: Zero-inflation equation ($zi$)}} \\[2pt]
EU-WB main effect & 1.27*** & {\scriptsize [0.80, 1.75]} & 0.38 & {\scriptsize [$-$0.27, 1.03]} \\
Trade intensity (w) $\times$ EU-WB & $-$0.32 & {\scriptsize [$-$1.26, 0.60]} & 0.05 & {\scriptsize [$-$0.89, 1.02]} \\
Trade intensity (b) $\times$ EU-WB & $-$0.11 & {\scriptsize [$-$0.85, 0.65]} & $-$0.02 & {\scriptsize [$-$0.88, 0.83]} \\
Spec.\ distance (w) $\times$ EU-WB & $-$0.27 & {\scriptsize [$-$0.77, 0.23]} & $-$0.16 & {\scriptsize [$-$0.78, 0.48]} \\
Spec.\ distance (b) $\times$ EU-WB & $-$0.03 & {\scriptsize [$-$0.53, 0.48]} & $-$0.35 & {\scriptsize [$-$1.04, 0.34]} \\
\bottomrule
\end{tabular}
\begin{tablenotes}
\scriptsize
\item Notes: Each block reports the pooled EU--WB convergence results from the FULL\_INT specification, stacked by prior structure. The default-prior block uses the flatter \texttt{brms} defaults for population-level coefficients; the harmonized block uses the regularizing project-wide prior structure. Both blocks report the same focal EU--WB terms from the mean and zero-inflation equations. Controls are identical across prior structures: lagged dependent variable, main effects for trade intensity and specialization distance (within and between), fiscal-gap, urban-gap, remittance-gap, inflation-gap, and financial-openness terms in the mean equation, plus year fixed effects and dyad random intercepts; the zero-inflation equation includes lagged dependent variable, trade, specialization distance, financial openness, year fixed effects, and dyad random intercepts. EU--WB = 1 if exactly one country in the pair is Western Balkan (MKD, SRB, MNE); the reference group is EU--EU pairs. WB--WB pairs are excluded. (w) = within-dyad effect; (b) = between-dyad effect. Stars indicate posterior credible-interval exclusion of zero: * 90\%, ** 95\%, *** 99\%. They are not p-values. Samples are identical across prior structures: short-run N = 7,025, dyads = 403, years = 20; long-run N = 6,468, dyads = 403, years = 19.
\end{tablenotes}
\end{threeparttable}
\end{table}

Tables~\ref{tab:wb_3way_harmonized} and~\ref{tab:wb_3way_default_priors} report the country-specific three-way WB analysis under the more regularized and less regularized prior structures, respectively. This country-specific evidence is more prior-sensitive than the pooled convergence table. Under the less regularized specification, the short-run zero-inflation gaps and some country-specific trade effects are larger, while under the more regularized specification many of these terms weaken or disappear. Even so, a narrower set of patterns remains visible across specifications: the negative short-run mean gap is concentrated in MKD, the short-run trade association over time is strongest for MKD and SRB, and the long-run negative structural effect over time is most visible for MNE. Because these country-specific models are estimated on a smaller and thinner sample than the pooled WB specification, several intervals remain wide. The country-specific zero-inflation results should therefore be read as exploratory, whereas the mean-equation results are the more robust part of the WB heterogeneity evidence.

\begin{table}[htbp]
\centering
\caption{Country heterogeneity results (baseline regularizing prior structure)}
\label{tab:wb_3way_harmonized}
\footnotesize
\setlength{\tabcolsep}{3.5pt}
\renewcommand{\arraystretch}{0.92}
\begin{threeparttable}
\begin{tabular}{lcccccc}
\toprule
& \multicolumn{3}{c}{Short-run (1.5--4.5y)} & \multicolumn{3}{c}{Long-run (4.5--8.5y)} \\
\cmidrule(lr){2-4} \cmidrule(lr){5-7}
Coefficient & MKD & SRB & MNE & MKD & SRB & MNE \\
\midrule
\multicolumn{7}{l}{\textit{Panel A: Main effects}} \\
EU--country main effect ($\mu$) & $-$0.37*** & $-$0.05 & 0.21 & 0.10 & 0.14 & 0.37* \\
                  & {\scriptsize [-0.59, -0.15]} & {\scriptsize [-0.36, 0.24]} & {\scriptsize [-0.13, 0.57]} & {\scriptsize [-0.16, 0.37]} & {\scriptsize [-0.23, 0.51]} & {\scriptsize [-0.06, 0.80]} \\
EU--country main effect ($zi$) & 0.39 & 0.90*** & 1.72*** & 0.23 & $-$0.21 & 0.57 \\
                  & {\scriptsize [-0.14, 0.94]} & {\scriptsize [0.32, 1.49]} & {\scriptsize [1.13, 2.30]} & {\scriptsize [-0.52, 0.98]} & {\scriptsize [-0.98, 0.56]} & {\scriptsize [-0.20, 1.35]} \\
\midrule
\multicolumn{7}{l}{\textit{Panel B: Trade interactions}} \\
EU--country main effect ($\mu$) & $-$0.47*** & $-$0.04 & 0.13 & 0.15 & 0.18 & 0.25 \\
                  & {\scriptsize [-0.77, -0.16]} & {\scriptsize [-0.33, 0.27]} & {\scriptsize [-0.38, 0.63]} & {\scriptsize [-0.17, 0.47]} & {\scriptsize [-0.21, 0.55]} & {\scriptsize [-0.29, 0.80]} \\
EU--country main effect ($zi$) & 0.39 & 0.96*** & 1.46*** & 0.21 & $-$0.14 & 0.50 \\
                  & {\scriptsize [-0.19, 0.98]} & {\scriptsize [0.37, 1.56]} & {\scriptsize [0.82, 2.13]} & {\scriptsize [-0.56, 1.00]} & {\scriptsize [-0.90, 0.65]} & {\scriptsize [-0.30, 1.28]} \\
(W) trade intensity $\times$ EU--country ($\mu$) & 0.57 & 0.84** & $-$0.03 & $-$0.22 & 0.18 & 0.09 \\
                  & {\scriptsize [-0.18, 1.32]} & {\scriptsize [0.05, 1.59]} & {\scriptsize [-1.02, 0.95]} & {\scriptsize [-1.02, 0.55]} & {\scriptsize [-0.50, 0.89]} & {\scriptsize [-0.80, 1.01]} \\
(W) trade intensity $\times$ EU--country ($zi$) & $-$0.10 & $-$0.26 & 0.02 & -0.00 & 0.07 & $-$0.01 \\
                  & {\scriptsize [-1.06, 0.86]} & {\scriptsize [-1.20, 0.69]} & {\scriptsize [-0.93, 0.98]} & {\scriptsize [-0.97, 0.97]} & {\scriptsize [-0.87, 1.01]} & {\scriptsize [-0.99, 0.97]} \\
(B) trade intensity $\times$ EU--country ($\mu$) & $-$0.24 & 0.14 & $-$0.13 & 0.16 & 0.17 & $-$0.24 \\
                  & {\scriptsize [-0.77, 0.32]} & {\scriptsize [-0.28, 0.59]} & {\scriptsize [-0.96, 0.70]} & {\scriptsize [-0.38, 0.72]} & {\scriptsize [-0.33, 0.64]} & {\scriptsize [-1.11, 0.62]} \\
(B) trade intensity $\times$ EU--country ($zi$) & $-$0.02 & 0.15 & $-$0.70 & $-$0.04 & 0.36 & $-$0.30 \\
                  & {\scriptsize [-0.86, 0.84]} & {\scriptsize [-0.65, 0.96]} & {\scriptsize [-1.58, 0.17]} & {\scriptsize [-0.96, 0.85]} & {\scriptsize [-0.57, 1.27]} & {\scriptsize [-1.20, 0.62]} \\
\midrule
\multicolumn{7}{l}{\textit{Panel C: Structural interactions}} \\
EU--country main effect ($\mu$) & $-$0.42*** & 0.00 & 0.14 & 0.15 & 0.15 & 0.34 \\
                  & {\scriptsize [-0.67, -0.17]} & {\scriptsize [-0.30, 0.32]} & {\scriptsize [-0.23, 0.52]} & {\scriptsize [-0.15, 0.44]} & {\scriptsize [-0.23, 0.53]} & {\scriptsize [-0.11, 0.79]} \\
EU--country main effect ($zi$) & 0.42 & 0.89*** & 1.64*** & 0.23 & $-$0.14 & 0.57 \\
                  & {\scriptsize [-0.15, 0.99]} & {\scriptsize [0.30, 1.50]} & {\scriptsize [1.04, 2.25]} & {\scriptsize [-0.54, 1.02]} & {\scriptsize [-0.93, 0.62]} & {\scriptsize [-0.23, 1.34]} \\
(W) specialization distance $\times$ EU--country ($\mu$) & $-$0.14 & 0.17 & 0.16 & $-$0.20 & $-$0.07 & $-$0.63** \\
                  & {\scriptsize [-0.41, 0.14]} & {\scriptsize [-0.37, 0.70]} & {\scriptsize [-0.36, 0.68]} & {\scriptsize [-0.53, 0.11]} & {\scriptsize [-0.49, 0.33]} & {\scriptsize [-1.13, -0.11]} \\
(W) specialization distance $\times$ EU--country ($zi$) & $-$0.10 & $-$0.38 & $-$0.18 & $-$0.05 & 0.18 & $-$0.38 \\
                  & {\scriptsize [-0.69, 0.50]} & {\scriptsize [-1.15, 0.40]} & {\scriptsize [-0.91, 0.55]} & {\scriptsize [-0.84, 0.73]} & {\scriptsize [-0.65, 1.05]} & {\scriptsize [-1.22, 0.43]} \\
(B) specialization distance $\times$ EU--country ($\mu$) & 0.10 & $-$0.06 & 0.32 & $-$0.15 & $-$0.04 & 0.08 \\
                  & {\scriptsize [-0.19, 0.41]} & {\scriptsize [-0.31, 0.21]} & {\scriptsize [-0.19, 0.80]} & {\scriptsize [-0.58, 0.28]} & {\scriptsize [-0.32, 0.24]} & {\scriptsize [-0.43, 0.58]} \\
(B) specialization distance $\times$ EU--country ($zi$) & $-$0.05 & 0.04 & 0.47 & $-$0.12 & $-$0.33 & 0.08 \\
                  & {\scriptsize [-0.73, 0.62]} & {\scriptsize [-0.52, 0.61]} & {\scriptsize [-0.33, 1.29]} & {\scriptsize [-0.95, 0.73]} & {\scriptsize [-1.10, 0.41]} & {\scriptsize [-0.83, 0.97]} \\
\bottomrule
\end{tabular}
\begin{tablenotes}
\scriptsize
\item Notes: Posterior means with 95\% credible intervals in brackets. The table reports country main effects and country-specific within/between interactions for trade intensity and specialization distance. All specifications are estimated with background controls: lagged dependent variable, main effects for trade intensity and specialization distance (within and between), fiscal-gap, urban-gap, remittance-gap, inflation-gap, and financial-openness terms in the mean equation, plus year fixed effects and dyad random intercepts; the zero-inflation equation includes lagged dependent variable, trade, specialization distance, financial openness, year fixed effects, and dyad random intercepts. EU-MKD, EU-SRB, EU-MNE = pairs involving one EU member and one WB country. Stars indicate posterior credible-interval exclusion of zero: * 90\%, ** 95\%, *** 99\%. They are not p-values. $\mu$ = mean equation; $zi$ = zero-inflation equation. (W) = deviation from a dyad's own historical mean; (B) = a dyad's long-run average relative to other dyads. Effective estimation sample: short-run N = 7,025, dyads = 403, EU-country dyads per WB country = 26 each; long-run N = 6,468, dyads = 403, EU-country dyads per WB country = 26 each.
\end{tablenotes}
\end{threeparttable}
\end{table}

\begin{table}[htbp]
\centering
\caption{Country heterogeneity results (prior structure with no regularization)}
\label{tab:wb_3way_default_priors}
\footnotesize
\setlength{\tabcolsep}{3.5pt}
\renewcommand{\arraystretch}{0.92}
\begin{threeparttable}
\begin{tabular}{lcccccc}
\toprule
& \multicolumn{3}{c}{Short-run (1.5--4.5y)} & \multicolumn{3}{c}{Long-run (4.5--8.5y)} \\
\cmidrule(lr){2-4} \cmidrule(lr){5-7}
Coefficient & MKD & SRB & MNE & MKD & SRB & MNE \\
\midrule
\multicolumn{7}{l}{\textit{Panel A: Main effects}} \\
EU--country main effect ($\mu$) & $-$0.48*** & $-$0.16 & $-$0.05 & 0.08 & 0.32 & 0.41 \\
                  & {\scriptsize [-0.71, -0.24]} & {\scriptsize [-0.48, 0.18]} & {\scriptsize [-0.45, 0.36]} & {\scriptsize [-0.26, 0.42]} & {\scriptsize [-0.14, 0.78]} & {\scriptsize [-0.10, 0.93]} \\
EU--country main effect ($zi$) & 0.74** & 1.45*** & 2.94*** & 0.56 & $-$0.65 & 1.40** \\
                  & {\scriptsize [0.07, 1.41]} & {\scriptsize [0.68, 2.22]} & {\scriptsize [2.19, 3.74]} & {\scriptsize [-0.81, 1.90]} & {\scriptsize [-2.05, 0.72]} & {\scriptsize [0.03, 2.78]} \\
\midrule
\multicolumn{7}{l}{\textit{Panel B: Trade interactions}} \\
EU--country main effect ($\mu$) & $-$0.86*** & $-$0.18 & $-$0.81 & 0.14 & 0.26 & 0.25 \\
                  & {\scriptsize [-1.31, -0.39]} & {\scriptsize [-0.54, 0.19]} & {\scriptsize [-2.45, 1.04]} & {\scriptsize [-0.40, 0.68]} & {\scriptsize [-0.25, 0.74]} & {\scriptsize [-1.83, 2.55]} \\
EU--country main effect ($zi$) & 1.41** & 1.93*** & 5.20** & 0.70 & 0.18 & $-$0.96 \\
                  & {\scriptsize [0.18, 2.70]} & {\scriptsize [1.00, 2.84]} & {\scriptsize [0.43, 10.03]} & {\scriptsize [-1.57, 3.01]} & {\scriptsize [-1.36, 1.80]} & {\scriptsize [-8.97, 6.76]} \\
(W) trade intensity $\times$ EU--country ($\mu$) & 1.49*** & 2.14*** & $-$0.63 & $-$0.68 & 0.33 & 0.69 \\
                  & {\scriptsize [0.35, 2.77]} & {\scriptsize [0.99, 3.36]} & {\scriptsize [-6.31, 4.99]} & {\scriptsize [-2.14, 0.74]} & {\scriptsize [-0.67, 1.40]} & {\scriptsize [-1.37, 3.24]} \\
(W) trade intensity $\times$ EU--country ($zi$) & $-$2.44 & $-$4.39** & 3.69 & $-$0.35 & 3.47 & $-$3.80 \\
                  & {\scriptsize [-7.63, 2.41]} & {\scriptsize [-8.02, -0.59]} & {\scriptsize [-5.26, 16.95]} & {\scriptsize [-8.02, 7.23]} & {\scriptsize [-2.53, 10.47]} & {\scriptsize [-16.17, 7.20]} \\
(B) trade intensity $\times$ EU--country ($\mu$) & $-$0.91* & 0.01 & $-$1.57 & 0.17 & $-$0.26 & $-$0.27 \\
                  & {\scriptsize [-1.81, 0.02]} & {\scriptsize [-0.62, 0.64]} & {\scriptsize [-4.75, 1.91]} & {\scriptsize [-0.98, 1.32]} & {\scriptsize [-1.01, 0.46]} & {\scriptsize [-4.34, 4.14]} \\
(B) trade intensity $\times$ EU--country ($zi$) & 1.61 & 1.38 & 4.47 & 0.37 & 3.26** & $-$4.53 \\
                  & {\scriptsize [-0.89, 4.16]} & {\scriptsize [-0.36, 3.16]} & {\scriptsize [-4.86, 13.79]} & {\scriptsize [-4.33, 5.22]} & {\scriptsize [0.26, 6.35]} & {\scriptsize [-19.77, 10.49]} \\
\midrule
\multicolumn{7}{l}{\textit{Panel C: Structural interactions}} \\
EU--country main effect ($\mu$) & $-$0.57*** & $-$0.10 & $-$0.15 & 0.06 & 0.31 & 0.32 \\
                  & {\scriptsize [-0.86, -0.29]} & {\scriptsize [-0.46, 0.25]} & {\scriptsize [-0.59, 0.28]} & {\scriptsize [-0.36, 0.48]} & {\scriptsize [-0.17, 0.79]} & {\scriptsize [-0.27, 0.94]} \\
EU--country main effect ($zi$) & 0.91** & 1.48*** & 2.88*** & 0.94 & $-$0.39 & 1.64* \\
                  & {\scriptsize [0.10, 1.72]} & {\scriptsize [0.64, 2.31]} & {\scriptsize [1.93, 3.85]} & {\scriptsize [-0.74, 2.58]} & {\scriptsize [-1.88, 1.10]} & {\scriptsize [-0.10, 3.33]} \\
(W) specialization distance $\times$ EU--country ($\mu$) & $-$0.17 & 0.27 & 0.22 & $-$0.23 & $-$0.07 & $-$0.83*** \\
                  & {\scriptsize [-0.45, 0.12]} & {\scriptsize [-0.35, 0.87]} & {\scriptsize [-0.37, 0.81]} & {\scriptsize [-0.57, 0.11]} & {\scriptsize [-0.53, 0.39]} & {\scriptsize [-1.38, -0.24]} \\
(W) specialization distance $\times$ EU--country ($zi$) & $-$0.37 & $-$1.45** & $-$0.57 & $-$0.09 & 0.39 & $-$1.61* \\
                  & {\scriptsize [-1.14, 0.44]} & {\scriptsize [-2.84, -0.03]} & {\scriptsize [-1.78, 0.67]} & {\scriptsize [-1.30, 1.22]} & {\scriptsize [-1.40, 2.27]} & {\scriptsize [-3.60, 0.18]} \\
(B) specialization distance $\times$ EU--country ($\mu$) & 0.20 & $-$0.04 & 0.45 & 0.02 & 0.05 & 0.23 \\
                  & {\scriptsize [-0.16, 0.58]} & {\scriptsize [-0.33, 0.25]} & {\scriptsize [-0.21, 1.13]} & {\scriptsize [-0.64, 0.66]} & {\scriptsize [-0.30, 0.42]} & {\scriptsize [-0.55, 0.99]} \\
(B) specialization distance $\times$ EU--country ($zi$) & $-$0.35 & $-$0.02 & 0.40 & $-$1.09 & $-$0.86 & $-$0.70 \\
                  & {\scriptsize [-1.43, 0.68]} & {\scriptsize [-0.81, 0.77]} & {\scriptsize [-1.39, 2.18]} & {\scriptsize [-3.35, 1.18]} & {\scriptsize [-2.23, 0.53]} & {\scriptsize [-3.55, 2.23]} \\
\bottomrule
\end{tabular}
\begin{tablenotes}
\scriptsize
\item Notes: Posterior means with 95\% credible intervals in brackets. The table reports country main effects and country-specific within/between interactions for trade intensity and specialization distance. All specifications are estimated with background controls: lagged dependent variable, main effects for trade intensity and specialization distance (within and between), fiscal-gap, urban-gap, remittance-gap, inflation-gap, and financial-openness terms in the mean equation, plus year fixed effects and dyad random intercepts; the zero-inflation equation includes lagged dependent variable, trade, specialization distance, financial openness, year fixed effects, and dyad random intercepts. EU-MKD, EU-SRB, EU-MNE = pairs involving one EU member and one WB country. Stars indicate posterior credible-interval exclusion of zero: * 90\%, ** 95\%, *** 99\%. They are not p-values. $\mu$ = mean equation; $zi$ = zero-inflation equation. (W) = deviation from a dyad's own historical mean; (B) = a dyad's long-run average relative to other dyads. Effective estimation sample: short-run N = 7,025, dyads = 403, EU-country dyads per WB country = 26 each; long-run N = 6,468, dyads = 403, EU-country dyads per WB country = 26 each.
\end{tablenotes}
\end{threeparttable}
\end{table}

\clearpage

\bibliography{bibliography}
\bibliographystyle{unsrt}

\end{document}